\documentclass[a4paper, oneside]{article}
\usepackage{a4wide}
\usepackage{graphicx,amssymb}
\usepackage{amsmath}
\usepackage{subfig}
\usepackage{float}
\usepackage[dutch,english]{babel}
\usepackage{color}
\usepackage{lineno}
\usepackage[export]{adjustbox}
\usepackage[column=0]{cellspace}
\usepackage{booktabs}
\usepackage{ragged2e}
\usepackage[colorlinks=true, allcolors=blue, hypertexnames=false]{hyperref}

\restylefloat{figure}
\graphicspath{{./Figures/}}

\begin{document}

\title{Cooperative transitions involving hydrophobic polyelectrolytes}
\author{James L. Martin Robinson and Willem K. Kegel} 
\date{\today}
\maketitle

\centering \large \textcolor{red}{\textbf{Please use the following link to access the published, citable, and \underline{corrected} version of this manuscript:}}
\\
\url{https://doi.org/10.1073/pnas.2211088120}
\\
\justifying \normalsize
\section*{Abstract}\label{abs}
Hydrophobic polyelectrolytes (HPE) can solubilize bilayer membranes, form micelles or can reversibly aggregate as a function of pH. The transitions are often remarkably sharp. We show that these cooperative transitions occur by a competition between two or more conformational states and can be explained within the framework of Monod - Wymann - Changeux (MWC) theory that was originally formulated for allosteric interactions. Here we focus on the pH-dependent destabilization and permeation of bilayer membranes by HPE. We formulate the general conditions that lead to sharp conformational transitions involving simple macromolecules mediated by concentration variations of molecular ligands. That opens up potential applications ranging from medicine to the development of switchable materials.
\section*{Introduction}\label{Intro}
Hydrophobic polyelectrolytes (HPE) are (bio) polymers that consist of hydrophobic as well as ionic (weak acids or bases) functional groups that are either part of the same side group or homogeneously distributed over the polymer chain (Table~\ref{table:polymer_structures}) \cite{Dobrynin1999, Thomas1994}. The main characteristic of the transitions involving HPE is their cooperative nature, or sharpness, as a function of $pH$.\\
In biology, small molecular ligands often bind to larger substrates, typically protein molecules, in a cooperative manner. These allosteric interactions can lead to sharp transitions between the unbound (without ligands) and bound states of a substrate \cite{Phillips2020}. The bound and unbound states are often related to some activity of the substrate, for example, low-molecular weight ligands can be activators or inhibitors of an enzyme \cite{Berg2011}. In that way, sharp activity switches, or ‘on’ and ‘off’ states, are possible as a function of the concentration of one or more ligands \cite{Phillips2013}. A textbook example of a cooperative binding transition is the allosteric binding of oxygen to hemoglobin in red blood cells. If there was no cooperativity (allostery) in the binding of oxygen molecules to the four binding sites of hemoglobin, the transition from unbound oxygen to complete saturation of all the four binding sites would occur over a relatively broad range in oxygen pressure. A model for the sharp transition over a relatively narrow range of oxygen pressure has been put forward by Monod, Wyman and Changeux \cite{Monod1965}, referred to as MWC theory. In that description, hemoglobin can be in (at least) two conformation states: one with low affinity for oxygen (the Tense or T state), and one with relatively high affinity (the Relaxed or R state). The last conformation is unfavorable at low oxygen concentration, but becomes favorable when several oxygen molecules bind. See the \hyperref[SI_theory]{\emph{SI:Theory}} section for a detailed description of MWC applied to hemoglobin. While it is likely that the oxygen binding sites of hemoglobin somehow interact, this is not a necessary assumption. In fact in the original MWC paper \cite{Monod1965}, no assumptions have been made regarding the interactions between oxygen binding sites. The only necessary requirement is that there are two conformational states, one of which is unfavorable at low oxygen pressure and becomes stable by binding more than one oxygen molecule. With that in mind, one may expect that cooperative transitions are not limited to complex substrates such as proteins, but may also occur in relatively simple substrates as long as there are well-defined conformation states, each with different binding affinity for ligands. This is what we will demonstrate in this work. A singular observation of cooperative binding by, likely, an MWC mechanism, has been reported on the binding of low-molecular weight ligands onto aggregates of modified cyclodextrins \cite{Petter1990}. An example of a mechanism other than MWC that can also cause sharp transitions in relatively simple substrates and ligands is by (weak) multivalent interactions, see, e.g. \cite{Martinez-Veracoechea2011, Dubacheva2019}. In these systems the transitions are driven by combinatorial entropy of multiple binding sites of both ligands and substrates (or receptors in the terminology in \cite{Martinez-Veracoechea2011}). Here we will focus on hydrophobic polyelectrolytes as relatively simple substrates that can show cooperative transitions driven by the concentration of potential determining ions. One conformation state of a hydrophobic polyelectrolyte chain is a hydrophobic state, where the ionizable groups in the polymer have a low affinity for potential determining ions, being protons in the case of basic side groups and hydroxyl ions for acidic groups. Another conformation can be an aqueous state which is unfavorable at low concentrations of potential determining ions but has a high affinity for these ions, and becomes stable upon binding several ions at once. Other conformations may also occur, such as HPE localized at the rim of a bilayer disk.\\ 
In the following we formulate MWC theory in terms of the properties of HPE and compare our model to the experimental behavior of hydrophobic polyelectrolytes in several guises: (1) micellization of HPE in the form of diblock copolymers, (2) the globule-coil transition in aqueous solutions, and (3) disk formation and permeation in lipid bilayer membranes. If it is possible to control transition $pH$ and the width of the $pH$ region where the transitions occur (the sharpness or degree of cooperativity), it will open up new routes to the development of artificial drug and gene delivery vehicles and cancer treatment, see for a review of $pH$-responsive tumor- targeted approaches \cite{Kanamala2016}, and \hyperref[SI_applications]{\emph{SI:Applications}} for more details. In broader perspective, the ability to switch between states with different optical, electronic, or other properties driven by (very) small concentration variations of relevant molecular ligands may aid in the development of new materials. 

\section*{Theory}\label{theory}
\begin{figure}[t]
\centering
\includegraphics[width=\textwidth]{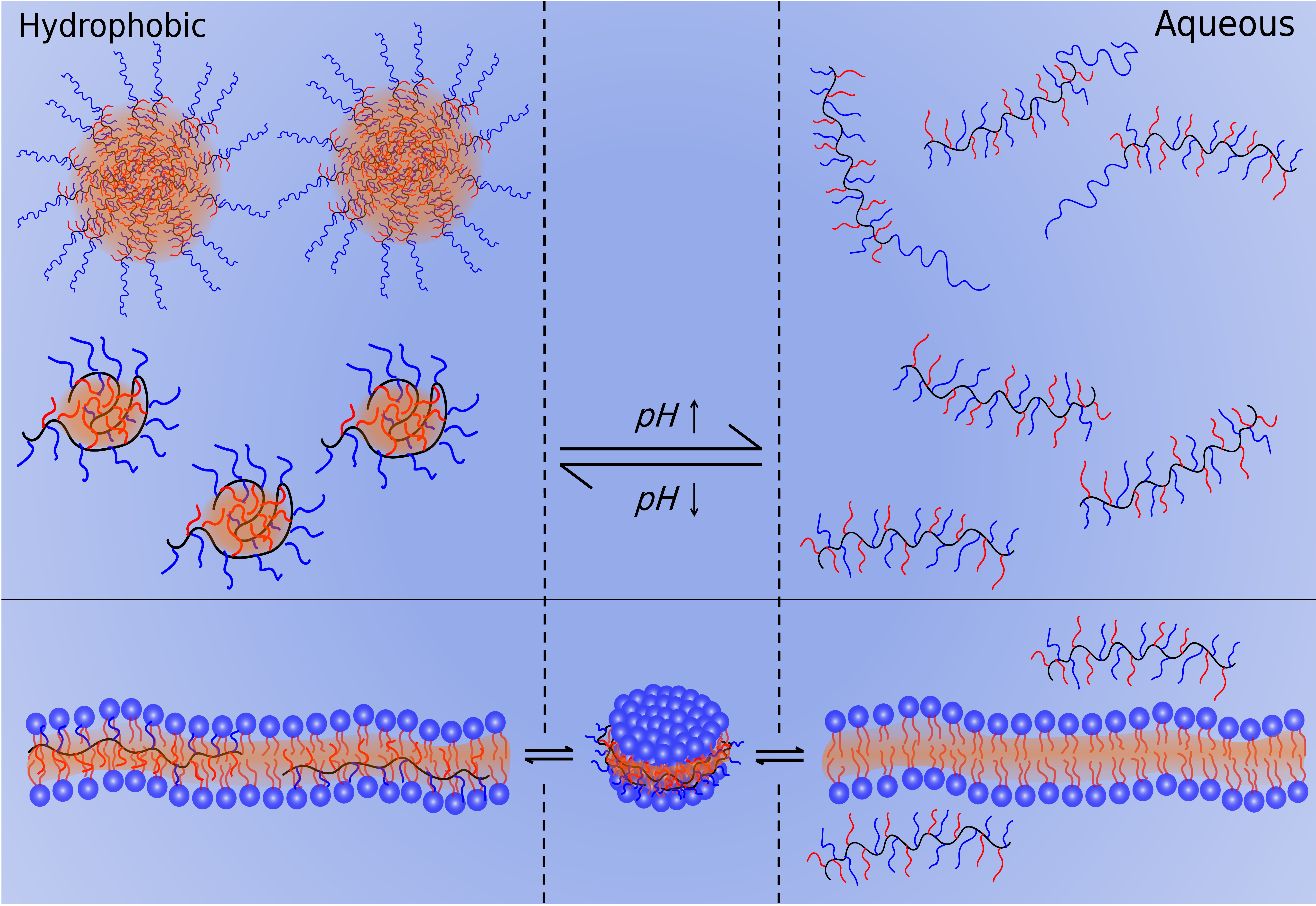}
\caption{Schematic illustration of a series of different possible states of an acidic hydrophobic polyelectrolytes in an aqueous solution. Changes in $pH$ will lead to changes in the conformation or environment of the chains. Top: micellization of a hydrophilic-hydrophobic polyelectrolyte diblock, middle: Coil to globule transition, bottom: Membrane permeation and solubilization. We may broadly partition them into "hydrophobic" and "aqueous" states, with a membrane nanodisk as an intermediate state. Red and blue polyelectrolyte side groups represent hydrophobic and hydrophilic (acidic) molecular moieties, respectively. Orange shading highlights the hydrophobic reservoirs that stabilise these hydrophobic side groups.}
\label{fig:partitioning}
\end{figure}

\begin{figure}[h]
	\centering
	\includegraphics*[width=1\linewidth]{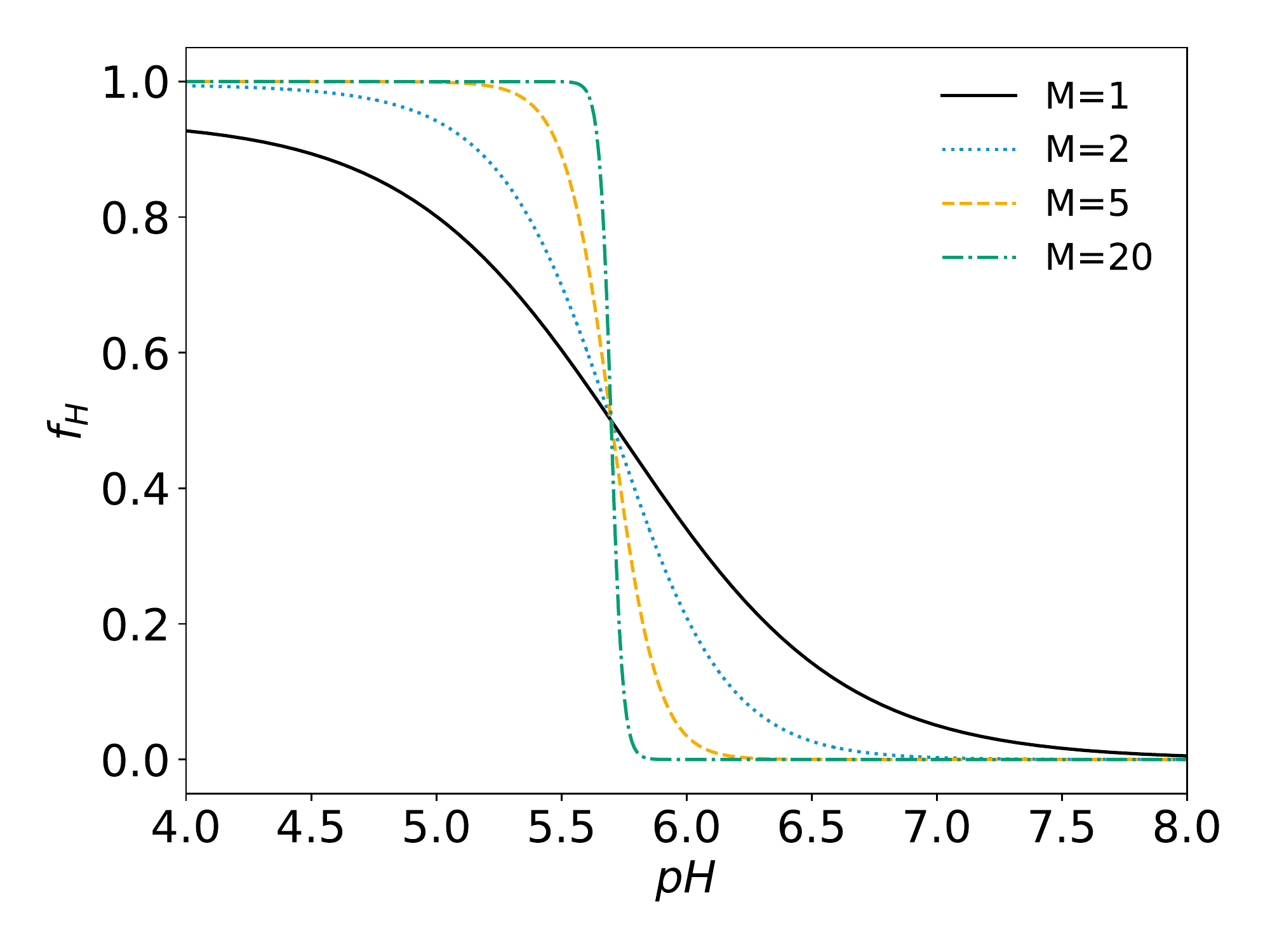}
	\caption{Plots of HPE conformational transitions from a hydrophobic to aqueous state, $f_H$ (Eq.~(\ref{eq:fraction})), for different degrees of cooperativity, $M$, in between 1 and 20. The number of ionizable acidic groups $M$ is equal to $M_H$, the number of hydrophobic groups. Further $pK_a = \text{4.5}$, and $G_H = M_H g_H$ where $\beta g_H = \text{2.82}$. Increasing $g_H$ shifts the transition to higher $pH$ for acidic groups.}
	\label{fig:fH}
\end{figure}

In the \hyperref[SI_theory]{\emph{SI:Theory}} we generalize MWC theory to the situation where substrates can have multiple states. In the case of HPE, we define a hydrophobic state as the ground state in which ionization is prohibitively unfavorable. The aqueous state is unfavorable in terms of interactions between the hydrophobic side groups of the polymers with water. On the other hand, in that state, ionization, or binding of potential determining ions (protons or hydroxyl ions), is favorable.  Transitions between these states are brought about by the changes in the concentrations of potential determining ions such as proton concentration, that is, $pH$. We can identify analogous transitions in the form of micellization of hydrophilic-hydrophobic polyelectrolyte diblocks, or the permeation and solubilization of membranes. We have schematically sketched these transitions in Fig.~\ref{fig:partitioning}.
\\    
The model we propose for the ionization transition is based on the Monod - Wyman - Changeux (MWC) theory for allosteric transitions \cite{Monod1965}, as described in the \hyperref[SI_theory]{\emph{SI:Theory}} section. We take the ionization of (weak) acid groups on the polymers as the binding of ligands in the form of hydroxyl ions \footnote{A proton hole is also an equivalent ligand choice.}. In the case of basic ionizable groups on the polymer, protons are the ligands. These (hydroxyl or proton) ligands play the role of oxygen in binding onto hemoglobin. We essentially neglect ionization in the hydrophobic state of the polymers, being similar to assuming a very low affinity for ligands in the form of ions. The statistical weight of a HPE in the aqueous state in the form of the coarse - grained grand partition function is given by, see Eq.~(\ref{ksiaqac}, \ref{ksiaqbs}):
\begin{linenomath}
\begin{equation}
\Xi_{aq} = \text{exp}(-\beta G_H)~(1 + 10^{X})^M. \label{ksiaq}
\end{equation}
\end{linenomath}
Here, $\beta$ is the inverse thermal energy and $G_H$ is the hydrophobic free energy ($>0$) penalty to transfer a polymer from its hydrophobic reference state to the situation where the polymer is in contact with water. It is worth noting that this penalty will also encapsulate other free energy changes when the environment of the chain changes, such as differences in conformational entropy. The hydrophobic reference state can be the collapsed globule state of a single polymer chain, or it can be the polymer as part of a large(r) aggregate with other chains. It can also be the interior of a micelle that can be formed if the hydrophobic polyelectrolyte is linked to a hydrophilic block, see Fig.~\ref{fig:partitioning} for illustration. For carboxyl groups with hydroxyl 'ligands' we have $X =pH- pK_a$, with $pK_a = - ^{10}\text{log}K_a$, $K_a$ being the dissociation constant of a (solvated) carboxyl group. For basic groups $X = pK_a' - pH$, with $pK_a' = - ^{10}\text{log}K_a'$, $K_a'$ being the dissociation constant for the conjugate acid of the basic group.\\
The value of $M$ should be seen as the maximum number of ionizable groups affected by the transition under the relevant conditions in terms of $pH$ and ionic strength. Coulomb interactions are expected to lead to values of $M$ that are significantly smaller than the number of ionizable groups on the polymer, see, e.g., \cite{Leyte1964}. The statistical weight of the reference hydrophobic state, is given by $\Xi_H \approx 1$. With the full grand partition function $\Xi = \Xi_{aq} + \Xi_H$ the fraction of polymer in the hydrophobic and aqueous state are given by
\begin{linenomath}
\begin{align}
f_H = \Xi_H/ \Xi &= \left(1 + \text{exp}(-\beta G_H)(1 + 10^X)^M \right)^{-1} \nonumber \\
&\text{and}~f_{aq} = 1 - f_H.
\label{eq:fraction}
\end{align}
\end{linenomath}
The fraction of ionized carboxyl or basic groups is given by, see Eq.~(\ref{SI_theta}) for derivation,
\begin{linenomath}
\begin{equation}
\theta = \frac{\langle N \rangle}{M} = \frac{10^X}{1 + 10^X}f_{aq}.\label{theta}
\end{equation}
\end{linenomath}
The value of $M$ in the equations above is a measure for the cooperativity ($M =1$ implies no cooperativity) and determines the steepness of the transition, in this case the $pH$ range where the transition takes place, from the aqueous to the hydrophobic state or vice versa. This behavior is illustrated in Fig.~\ref{fig:fH} for an HPE with equal numbers of hydrophobic and ionizable groups.
The $pH$ at which the transition takes place is in this model determined by the values of $pK_a$ or $pK_a'$ and $G_H$. The last quantity is expected to be an increasing function of the number of hydrophobic (side) groups of polymers that contain separate hydrophobic and ionizable groups, which has indeed been assumed in Fig.~\ref{fig:fH}. In the case of polymers that contain more than one type of hydrophobic (side) group, $G_H$ is expected to be a linear combination of the fraction of hydrophobic groups and their separate hydrophobic free energy contribution.\\
The equations above assume uncorrelated ionizable groups, which is reflected in a single value for the ionization constant. However, as mentioned above, Coulomb interactions will inevitably lead to a spread of this constant. We may interpret the equations, therefore, as capturing only the differences between the state of the polymer around the transition and not as an absolute description of the ionization state of the chain. In other words, around the transitions, we approximate correlations by assuming that at most only a single chargeable group is ionized over a length of approximately the Bjerrum length ($\approx$~0.72 nm in water).  We emphasize that this approximation holds in this case as further ionization beyond the point where the HPE are overwhelmingly in the aqueous state does not influence the transition. Including correlations explicitly in the theory at this point is a significant challenge, in particular as intermediate states of the HPE (besides hydrophobic, aqueous, disk) are expected.\\ 
As mentioned in the \hyperref[Intro]{\emph{Introduction}} section, when mixed with lipid bilayer membranes in the form of unilamellar or multilamellar vesicles, several hydrophobic polyelectrolytes spontaneously induce the formation of disks of roughly 10 nm in diameter throughout a well-defined $pH$ range, see \cite{Thomas1992, Thomas1994, Scheidelaar2016}. The polymers are thought to adsorb at the rims or the disks, see the bottom row of Fig.~\ref{fig:partitioning}, where the hydrophobic parts of the polymers stick into the hydrophobic interbilayer spacing and the ionic groups preferably orient towards the outside, thereby maximizing contact with water. More complex hydrophobic polyelectrolytes \cite{Vial2007, Vial2009, Yessine2004} and peptides \cite{Takechi2012, Copolovici2014} also destabilize membranes, probably by other mechanisms. So-called 'membrane scaffold proteins' (MSP) \cite{Denisov2016, Denisov2004, Bayburt2002, Morgan2011} can also form disks but only after treatment with detergents. Mixtures of two lipids with different size, or lipid with surfactant can also form disk-shaped aggregates on the order of tens of nanometers in size. These aggregates, where the smaller lipids or surfactant are located near the rims of the disks are referred to as 'bicelles' \cite{Durr2013} and can form upon appropriately mixing the components. \\
We define a third conformational state of the polymer when adsorbed onto disks. In the disk state, we assume that the hydrophobic parts of the polymer still pay a penalty for being at a hydrophobic - aqueous interface but that this penalty should be significantly smaller than for being fully in the aqueous state. As explained in the \hyperref[SI_theory]{\emph{SI:Theory}}, on average, less chargeable groups may get ionized compared to the situation where the polymers are fully dissolved in the aqueous state.  Considering only carboxyl ionic groups, the resulting grand canonical weight of a HPE bound onto a bilayer disk is given by
\begin{linenomath}
\begin{equation}
\Xi_{D} = \text{exp}(-\beta G_{HD})~(1 + 10^{pH - pK_a})^{M_{D}}, \label{ksid}
\end{equation}
\end{linenomath}
where $G_{HD}$ stands for the hydrophobic penalty of an HPE chain when adsorbed onto a disk which includes the formation free energy of the disk (per HPE chain) and $M_D$ is the number of chargeable groups on the HPE in the disk state. The value of $G_{HD}$ is expected to be a fraction of the value of $G_H$ in Eq.~(\ref{ksiaq}). $(M - M_{D})/M$ should be seen as the fraction of time chargeable groups spend inside or close to the hydrophobic bilayer region. This is analogous to adding an additional term in Eq.~(\ref{ksid}) of the form $(1 + 10^{pH - pK_{a}^*})^{M - M_{D}}$, with $pK_{a}^{*} >> pK_a$. We will see later that for the particular hydrophobic polyelectrolytes we consider, the fraction of nonionized groups can be very small ($<$0.1). Alternatively, there may be an additional structural contribution to that fraction based on the architecture of the polymers.
Moreover, the $pK_a$ of the available ionized groups may be different to the one in the unconstrained aqueous form due to electrostatic repulsion with the lipid head groups. Note that if $M_D = M$, disks are always stable with respect to the aqueous state as $G_{HD} < G_H$ and therefore the statistical weight $\Xi_D > \Xi_{aq}$, see Eqs. (\ref{ksiaq},\ref{ksid}).
The ionized fraction now reads, see Eq.~(\ref{SI_theta}),
\begin{linenomath}
\begin{multline}
	\theta =  \frac{10^{X}}{\Xi} \left( \frac{M_{D}}{M} \text{exp} (- \beta  G_{HD}) \left(1+ 10^{X} \right)^{M_{D} - 1}\right.
	\\+ \left. \text{exp} (-\beta G_H) \left(1 + 10^{X} \right)^{M - 1} \right) \label{thetaD}.~~\text{~(presence of disks)}
\end{multline}
\end{linenomath}
In Eq.~(\ref{thetaD}), $\Xi = \Xi_{aq} +\Xi_H + \Xi_{D}$. 
\section*{Comparison to experiments}
\subsection*{Diblock micelles}\label{micelles_section}
\begin{figure}[h]
	\centering
	\subfloat[\centering Fitted $f_{aq}$ (Eq.~(\ref{eq:fraction})) curves and $\theta$ (for 40:40 ratio, black rhombuses) curves (Eq.~(\ref{theta})). Note the strong correlation between the ionized and aqueous fraction in the middle curve (40:40 data). Global parameters: $M=11$, $g_{DBA}=11.2k_BT$ and $g_{DPA}=9.0k_BT$. $pK_{a}'=10.1$ from \cite{Li2016}.]{{\includegraphics[width=0.5\linewidth]{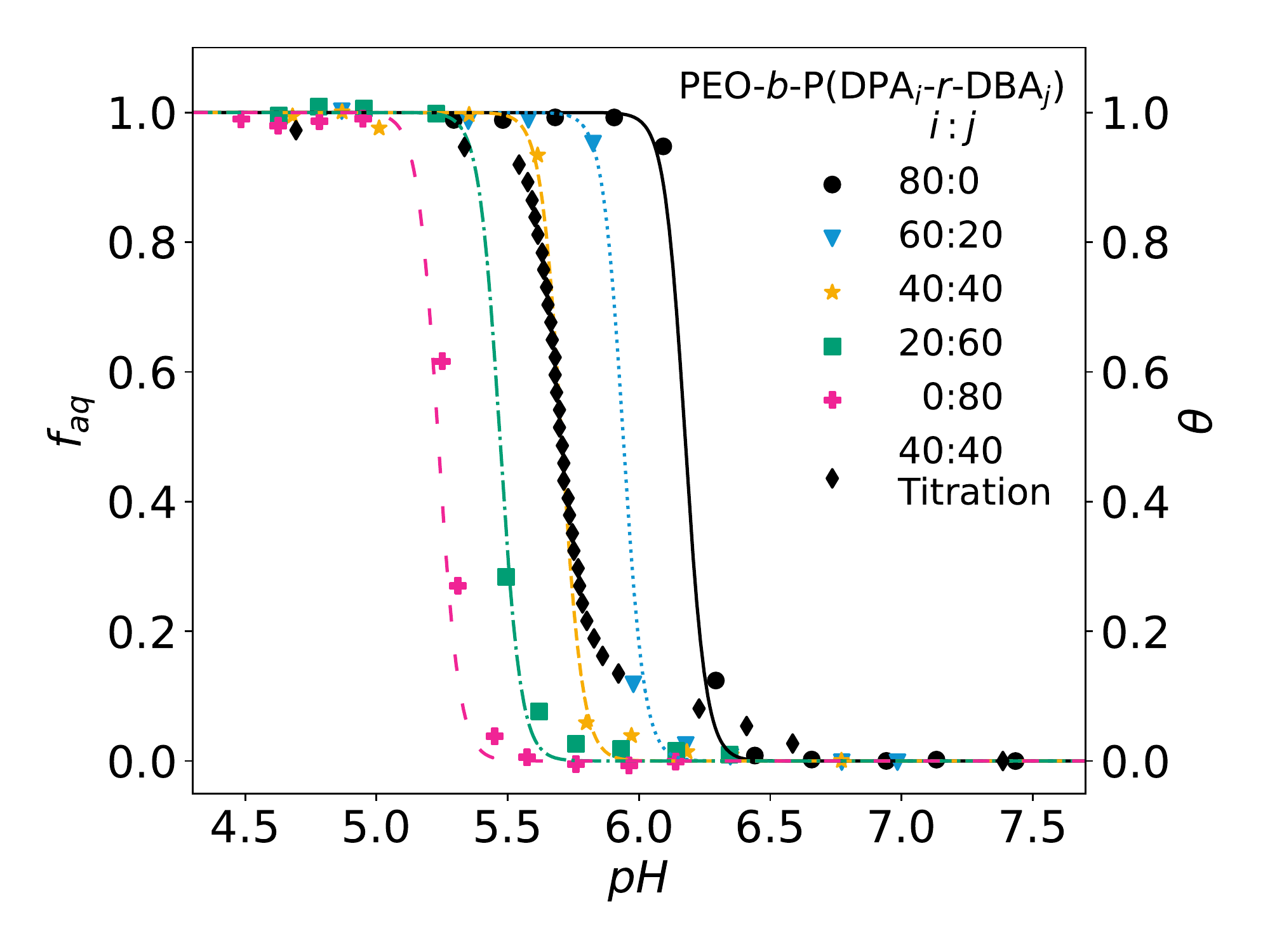} }}%
	\subfloat[\centering Transition $pH$ trend with respect to DBA mole fraction in the HPE block ($x$), fit with Eq.~(\ref{pHmic}) using $pK_{a}'=10.1$ from \cite{Li2016}.]{{\includegraphics[width=0.5\linewidth]{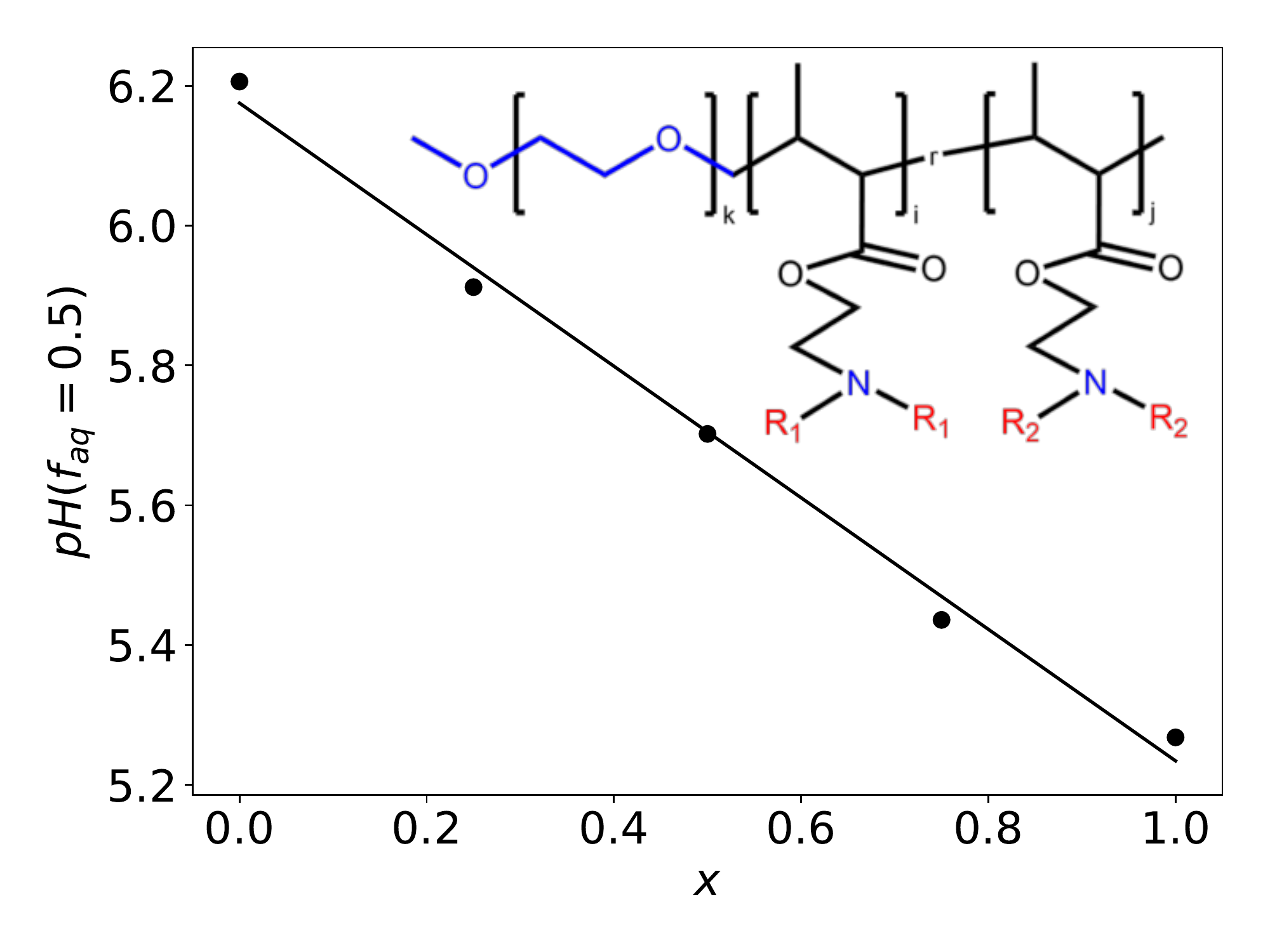} }}%
	\caption{Analysis of PEO-b-PDPA-r-PDBA micellization data from Ref. \cite{Ma2014}. (a) Fluorescent and titration data. (b) Transition $pH$ trend. Inset: Chemical structure of the polymer.}
	\label{comp_fig}
\end{figure}

\begin{figure}[h!]
    \centering
    \subfloat[\centering Fit of the titration data using Eq.~(\ref{theta}) for different lengths ($i$) of DPA blocks in PEO-b-PDPA diblocks. Local parameters: $M=1.8,3.1,5.5,8.4,45.3$ for the 5, 10, 20, 60 and 100 lengths respectively. Values of $G_H$ are calculated from Eq.~(\ref{pHmic2}) with global paremeters: $b=0.60$ , $g_H=0.9k_BT$. $pK_a'=10.1$ from \cite{Li2016}. Inset: Fitted $M$ against DPA block length ($i$) trend.]{{\includegraphics[width=0.5\linewidth]{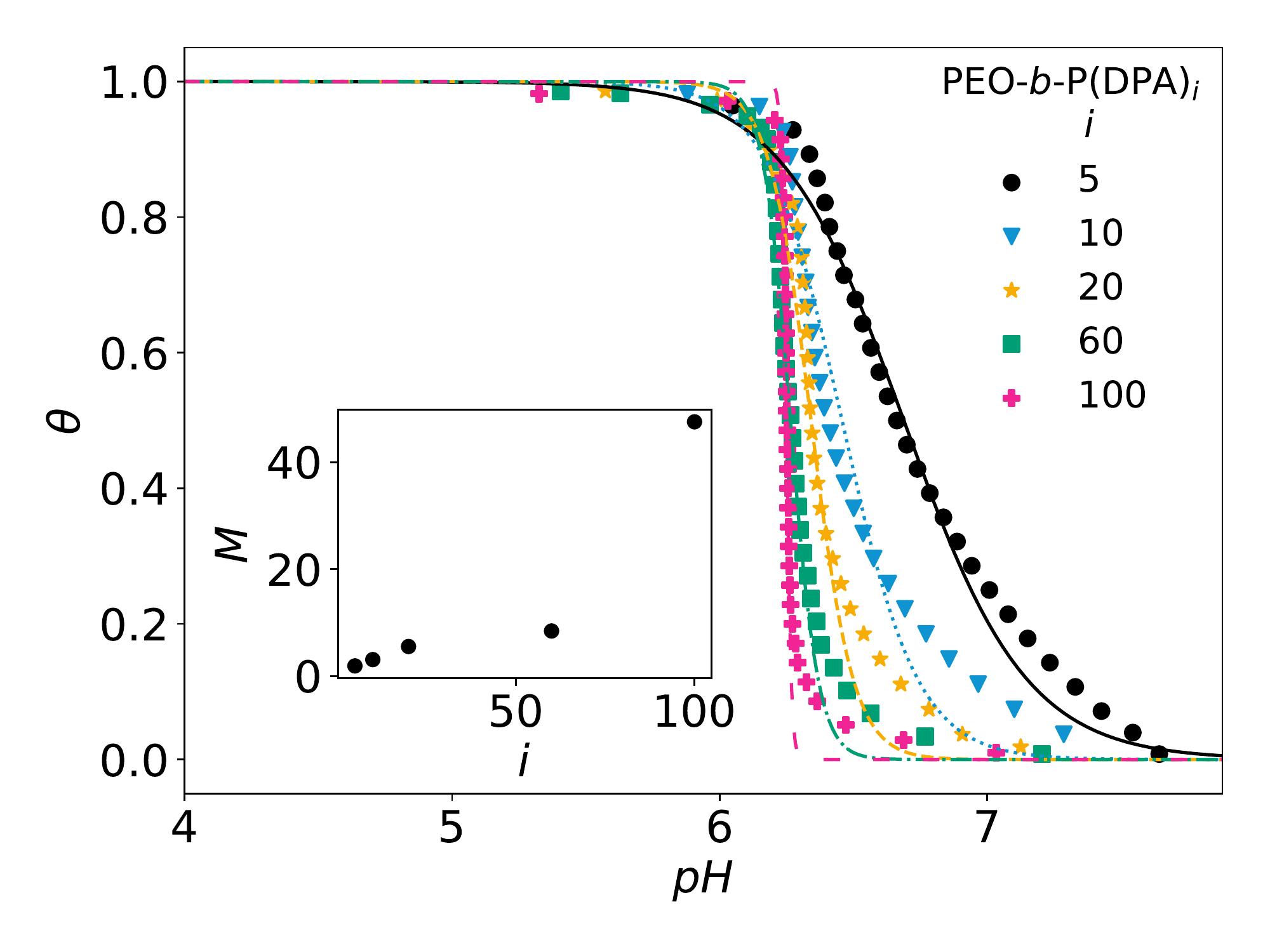} }}
    \subfloat[\centering Fit, using Eq.~(\ref{pHmic2}), of the transition $pH$ with respect to DPA block length ($i$). $M=i$ was set in Eq.~(\ref{pHmic2}). Global parameters: $b=0.60$ , $g_H=0.9k_BT$. $pK_a'=10.1$ from \cite{Li2016}.]{{\includegraphics[width=0.5\linewidth]{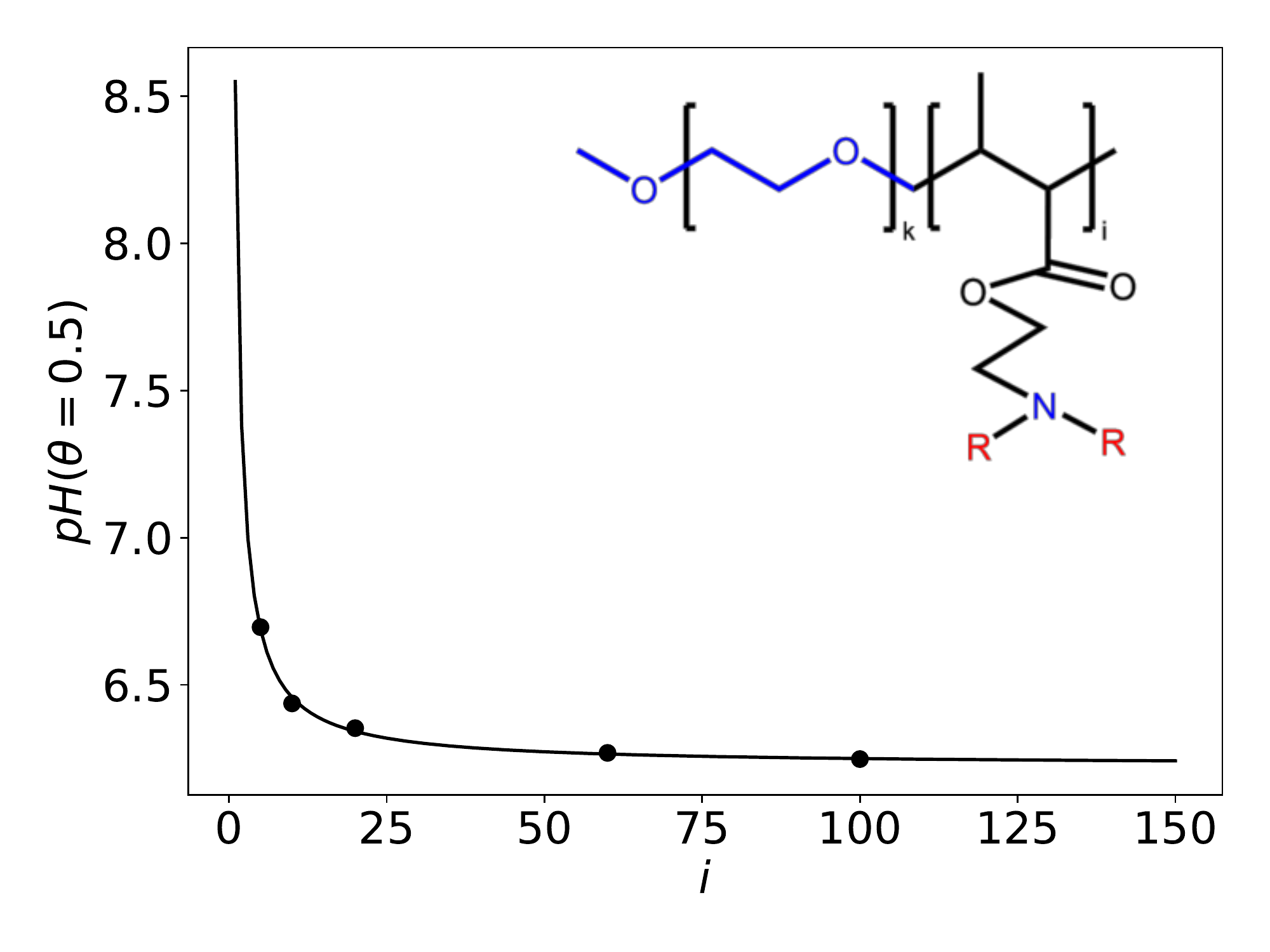} }}%
    \caption{Analysis of the effect of PDPA block length on PEO-b-PDPA titration data from Ref. \cite{Li2016}. (a) Titration data. (b) transition $pH$ trend. Inset: Chemical structure of the polymer.}
    \label{Gao_fig}
\end{figure}

An excellent example of a well-defined system which incorporates HPEs are the family of polymers investigated by Gao et al. \cite{Zhou2011,Ma2014,Li2016, Li2016_2}. They synthesize diblock copolymers where one of the blocks is a cationic hydrophobic polyelectrolyte based on tertiary amines. Reference \cite{Ma2014} includes fluorescent quenching data for a series of poly(ethylene oxide)-b-poly(2-(dipropylamino) ethyl methacrylate)-r-poly(2-(dibutylamino) ethyl methacrylate) (PEO-b-PDPA-r-PDBA) random copolymers (see inset in Fig.~\ref{comp_fig}b), the fluorescence correlating with the dissolution of the micelles at low $pH$ (see top row of Fig.~\ref{fig:partitioning} for a schematic of the transition). The series of polymers have different DPA to DBA ratios ($i:j$) which affect their transition $pH$. From Fig.~\ref{comp_fig}, reproduced from their work, we can observe sharp transitions for a series of different compositions. The $pH$ range in which the transition occurs is over approximately 0.2 $pH$ units, which is remarkably sharp and points to a highly cooperative transition (indeed $M \approx$ 11, see later). To compare, a typical monomer analog of an HPE, butyric acid in a demixed system of octanol and water, undergoes a transition from mainly soluble in octanol to mainly soluble in water over a $pH$ range as broad as roughly 4 $pH$ units \cite{Rayne2016}, which also is obvious in Fig.~\ref{fig:fH} for $M = 1$. It can also be seen in Fig.~\ref{comp_fig}a that upon increasing the fraction of the most hydrophobic group in the polymers, the transition from an aqueous coil to a micelle (hydrophobic) state occurs at decreasing $pH$.
The aliphatic chains on this amine are varied to impart different degrees of hydrophobicity therefore allowing for control over the transition $pH$. Due to the hydrophilic PEO block there is a drive to form well-defined micelles, which we denote as the hydrophobic state in this system. In other words, the hydrophilic blocks prevent the formation of macroscopic aggregates and stabilizes the HPE in their hydrophobic state in the cores of well-defined micelles. These will form at a $pH$ high enough for the amines to become deprotonated in the core of the micelle. Conversely the micelles fall apart as the $pH$ is lowered and the ionization of the HPE blocks is favored leading to an increased solubility in the aqueous solution.
\\We compare our model in the form of Eqs.~(\ref{eq:fraction}, \ref{theta}) to the experiments in Fig.~\ref{comp_fig}a. The structure of the hydrophobic polyelectrolyte indicates that the number of ionizable acidic groups $M$ is equal to $M_H$, the number of hydrophobic groups ($M_H = M$). As explained in the \hyperref[theory]{\emph{Theory}} section we can describe the hydrophobic penalty, $G_H$, as a linear combination of the hydrophobic penalty for the two types of monomers via $G_H = M f(g_{DBA}, g_{DPA}, x)$. Where $f(g_{DBA}, g_{DPA}, x) = x g_{DBA} + (1-x) g_{DPA}$. $g_{DBA}$, $g_{DPA}$ are the hydrophobic free energy contributions per butyl and propyl monomer respectively, and $x$ is the mole fraction of DBA side groups in the hydrophobic polyelectrolyte. An important property of the diblock copolymer micelles is their transition $pH$, which we define as $\Xi_{H}=\Xi_{aq}(=1)$ leading to $f_{aq}=0.5$ and the following expression:
\begin{linenomath}
\begin{equation}
pH_{\text{micellization}} = pK_a' - \text{0.4343}\beta f(g_{DBA}, g_{DPA}, x) \label{pHmic}.
\end{equation}
\end{linenomath}
Here we further used that at the transition, $10^{pKa' - pH} >>1$. The numerical factor 0.4343 $\approx 1/\text{ln10}$. A $pK_a'$ of 10.1 was taken \cite{Li2016}. 
There is an excellent agreement between this linear approximation and the experimental data, as shown in Fig.~\ref{comp_fig}b. Transition $pH$ for different compositions can be easily calculated. Therefore, co-polymerizing monomers with different hydrophobicities is an excellent method to target specific transition $pH$ values, as also pointed out in \cite{Ma2014}.
Introducing the extracted values of $G_H$ into Eq.~(\ref{eq:fraction}) and using an average value of $M$ calculated from a preliminary free parameter fit leads to the theoretical curves shown in Fig.~\ref{comp_fig}a. Again, the general shape of the data is adequately described.\\
Equation (\ref{theta}) describes the relationship between $f_H$ (and $f_{aq}$) and the ionization state of the polymer. In the language of MWC theory, protonation is similar to the binding of ligands which in turn drives the conformation transition from a micelle (unbound to protons) to an aqueous (bound to at least several protons) state. Therefore in MWC theory ionization and aqueous fraction are predicted to be strongly correlated. This correlation has been illustrated for oxygen binding onto hemoglobin in Fig.~\ref{fig:Hb}. In terms of our version of the theory, the factor relating $f_{aq}$ and $\theta$, \(\frac{10^{pK_a' - pH}}{1 + 10^{pK_a' - pH}}\) (for a basic polyelectrolyte), will be close to unity at $pH$ values a couple of units below the $pK_a'$, where most transitions take place. Therefore we expect a similar transition both in sharpness and transition $pH$ to be present for both titration and fluorescence data. Comparison of titration and fluorescence data from \cite{Ma2014} in Fig.~\ref{comp_fig}a indeed shows good correspondence between both transitions. There is a strong correlation between the ionization state of the polymer and the conformational state of the chain, integral to the model we propose. The overlap of both of the transitions is however not perfect. There is a clear increase in ionization fraction of the polymer before the micelle dissolves. This behavior may arise due to outer groups on the HPE block becoming partially charged before the groups in the core.  
\\The value of $M$ refers to the number of groups which change their ionization state during the transition. From Fig.~\ref{comp_fig} it is clear that the polymer goes from mainly ionized to mainly deionized during the transition. If ionization (protonation) was uncorrelated, which probably is the most severe approximation that we used, and in the absence of other broadening effects, the fitted value of $M$ would be equal to the number of ionizable groups on the polymers, in this case 80. The average fitted value of $M$ is however around 11. Correlations due to Coulomb interactions will lead to a significant fraction of ionizable groups that remain uncharged around the $pH$ where most micelles have dissolved. Other reasons for this effective widening of the transition are that roughly 20\% of the polymer is already ionized before the micelles start dissolving. The two- state model relies on the assumption that there are solely two distinct environments for the chain. In reality there will most likely be intermediate states between the two extremes of the micelle and the single chain. This will widen the transition. In the model proposed this would correspond to states with slightly different $G_H$ values. A system with one intermediate state is investigated in the \hyperref[SI_broadening]{\emph{SI: Influence of intermediate states on transition broadening}} section. There also might be experimental reasons for some broadening, but these are expected to be small, if only because hardly any hysteresis has been observed in these systems \cite{Li2016}. In other words, the ionization of the micelle system is identical during the formation and dissolution of the micelles.      
\\Next we test the prediction from Eqs.~(\ref{eq:fraction}, \ref{theta}) that the coil - micelle transition becomes more cooperative (sharper) upon an increasing number of ionizable groups $M$ or chain length. In Refs. \cite{Li2016, Li2018} Gao et al. present a system of diblock poly(ethylene oxide)-b-poly(2-(dibutylamino) ethyl methacrylate) (PEO-b-PDBA) polymers containing a hydrophilic block and a hydrophobic polyelectrolyte whose chain length ($i$) is varied (see inset in Fig.~\ref{Gao_fig}b). The data, in the form of titrations, is presented in Fig.~\ref{Gao_fig}a. This is a similar system to the one described above however as the composition of the side groups is constant here, one would expect from Eq.~(\ref{pHmic}) that the transition $pH$ is independent of chain length. That clearly is not the case, as can be seen in Fig.~\ref{Gao_fig}b: the transition $pH$, defined as the $pH$ where $\theta = \text{0.5}$, decreases over roughly 0.5 $pH$ units when the number of monomers per chain increased from 5 to 100, the effect being largest for the shorter polymers.  We postulate that this systematic change with length is caused by finite-size effects due to the hydrophobic groups that are close to the hydrophobic - aqueous interface. The groups on the HPE blocks neighboring the hydrophilic blocks experience a higher polarity environment than the groups that are further away from the junction and are immersed in the center of the core of the micelle. This can be accounted for by writing $G_H = g_H(M - b)$, where the value of $b$ represents the portion of the polymer that sits close to the hydrophilic-hydrophobic junction or micelle interface. The value of $b$ is expected to be smaller than unity, that is, a fraction of a monomer unit. So we modify Eq.~(\ref{pHmic}) into
\begin{linenomath}
\begin{equation}
pH_{\text{micellization}} = pK_a' - \text{0.4343}\beta g_H \frac{M - b}{M} \label{pHmic2}.
\end{equation}
\end{linenomath}
This approach leads to a good description of the transition $pH$ (Fig.~\ref{Gao_fig}b) and incorporating this expression for $G_H$ into Eq.~(\ref{theta}) allows us to fit the transitions adequately (Fig.~\ref{Gao_fig}a). In this case a separate effective value of $M$ is fit to each curve.  
The values of $M$ that follow from the fits of the experimental data in Fig.~\ref{Gao_fig} to Eq.~(\ref{theta}) again are significantly smaller than the number of ionizable groups, being consistent with the analysis of the data in Fig.~\ref{comp_fig}. However, the trend illustrated in the inset of Fig.~\ref{Gao_fig}a is clear: already with a chain length as short as 5 monomers, the transition is cooperative with $M \approx$ 1.8 (no cooperativity corresponds to $M=1$). At the longest chain of 100 monomers, we find $M \approx 45$. While it is unclear at this point what causes the large increase in effective $M$ value in between chain lengths of 60 and 100 monomers, the increased level of cooperativity as a function of chain length is consistent with the trend predicted by theory. The significant cooperativity of the transitions on the basis of Fig.~\ref{Gao_fig}a was also concluded in \cite{Li2016}, based on the Hill coefficient \cite{Hill1910}. As Hill isotherms are empirical and can reflect many mechanisms \cite{Hill1910}, the added value of the analysis here is that the cooperative nature of the transition is shown to be coupled to the conformations of the polymers via an MWC-like mechanism. All of this combined constitutes the most complete proof of an MWC-like transition in relatively simple (macro) molecules that is currently available from existing data in the literature. That, in turn, leads to the question whether all transformations induced by $pH$ in HPE are consistent with the MWC model. As will be illustrated below for the single-chain globule-coil transition in HPE, that is not always the case, at least not convincingly. \\

\subsection*{Coil-Globule transition}
\begin{figure}[h!]
    \centering
    \subfloat[\centering PMAA: Fits, using Eq.~(\ref{eq:fraction}), of rotation correlation data (black) (parameters: $M=1.2$, $g_H=2.8k_BT$) and dynamic light scattering (DLS) data (cyan) (parameters: $M=2.9$, $g_H=2.9k_BT$). $pK_a=4.5$ is assumed. Titration ($\theta$) data is shown in orange.]{{\includegraphics[width=0.5\linewidth]{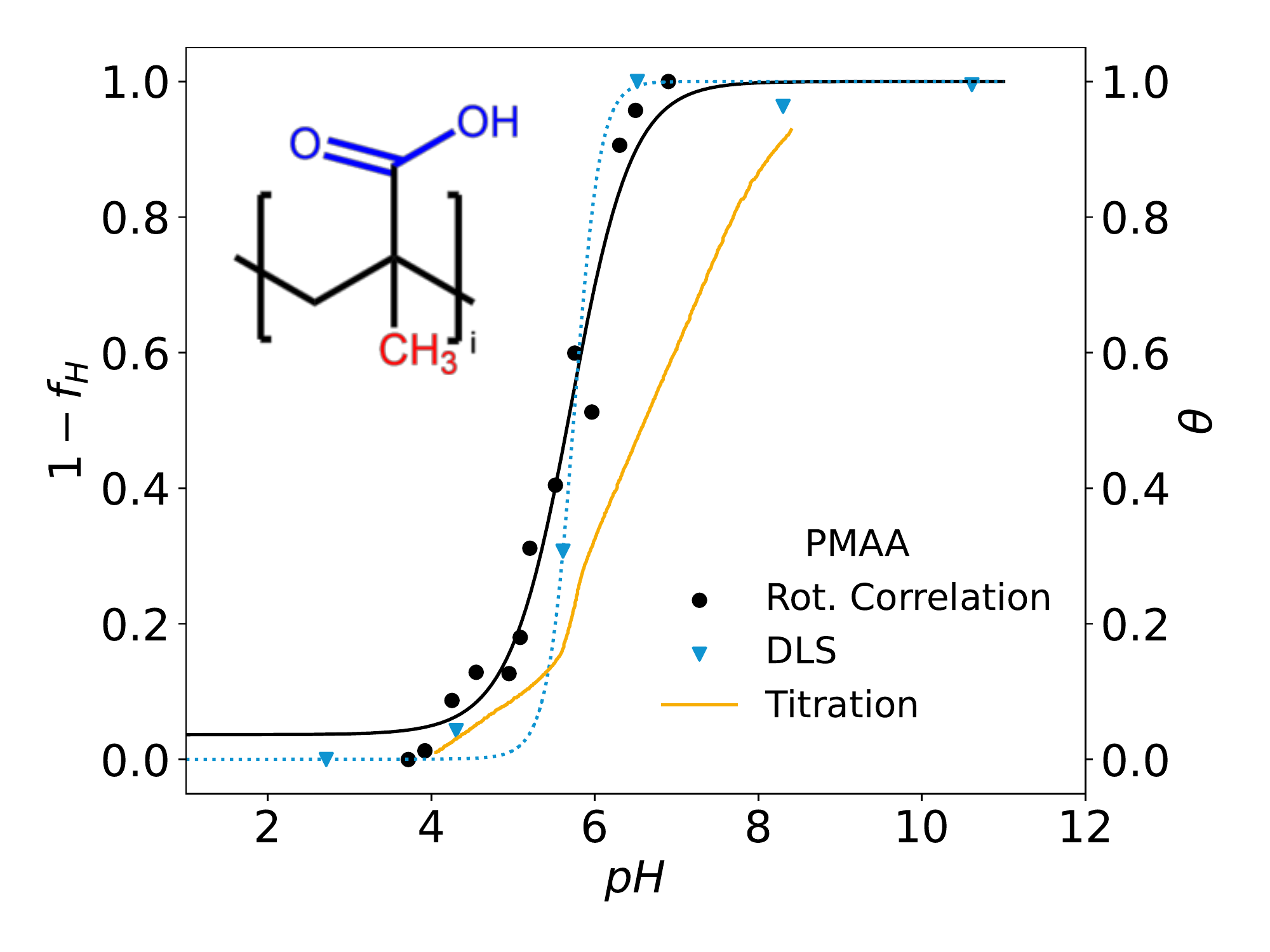} }}
    \subfloat[\centering PEAA: Fits, using Eq.~(\ref{eq:fraction}), of pyrene probe fluorescence data (black) (parameters: $M=3$, $g_H=3.3k_BT$) and dynamic light scattering (DLS) data (cyan) (parameters: $M=6$, $g_H=3.6k_BT$). $pK_a=4.5$ is assumed. Titration ($\theta$) data is shown as black stars.]{{\includegraphics[width=0.5\linewidth]{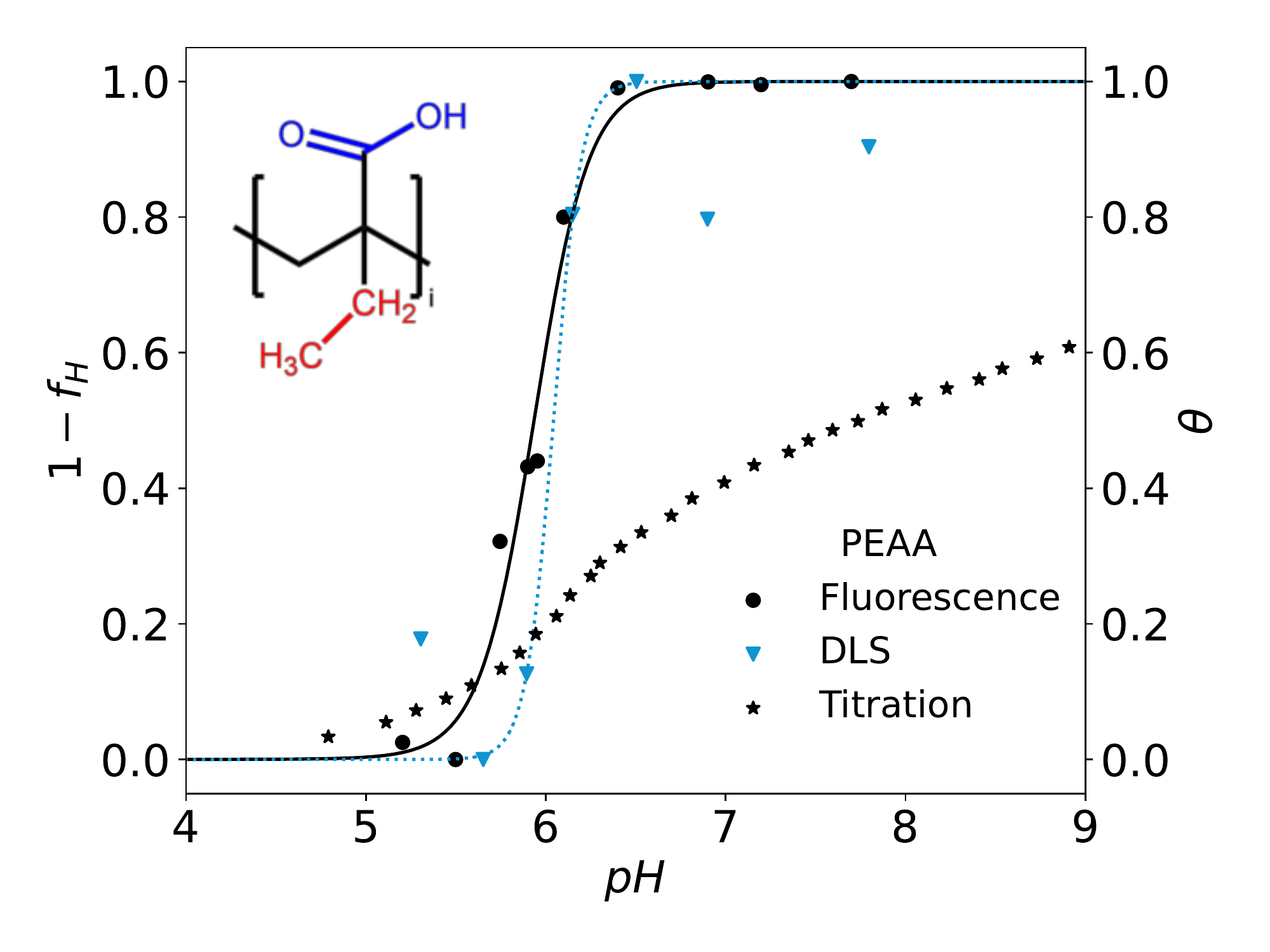} }}%
    \caption{Coil-globule transitions for (a) PMAA and (b) PEAA as a response to solution $pH$ changes. This data was obtained from Ref. \cite{RuizPerez2008} and \cite{Leyte1964}, and  Ref. \cite{Eum1989} and \cite{Thomas1995} respectively. Insets: chemical structure of the polymers.}
    \label{Coil_Glob}
\end{figure}

The coil to globule transition is, a priori (and perhaps naively), the most basic example of a two-state system we may consider (see middle row of Fig.~\ref{fig:partitioning}). However, at least in (slightly) hydrophobic polyelectrolytes, the debate still continues after the early work of Mandel et al. \cite{Mandel1967} and Koenig et al. \cite{Koenig1969} in the late 1960s. A more recent attempt to pin down the nature of the coil-globule transition in poly(methacrylic acid) (PMAA) was reported by Ruiz et al. \cite{RuizPerez2008}. They effectively look at the transition at different length scales by comparing rotational correlations (monomer scale) and hydrodynamic radii (full polymer scale).
\\
We combine the data from Ref.~\cite{RuizPerez2008} with those on poly(ethylacrylic acid) (PEAA) in \cite{Eum1989} in Fig.~\ref{Coil_Glob}. There we also add the measured ionization fractions $\theta$ obtained by titration from Ref.~\cite{Leyte1964} (Fig.~\ref{Coil_Glob}a) and Ref.~\cite{Thomas1995} (Fig.~\ref{Coil_Glob}b). To apply the analysis described in the previous section to the experimental data, those were scaled and translated into a hydrophobic fraction. The rotation correlation spectroscopy data, which probes local viscosity in a molecule, can be directly translated into a hydrophobic fraction by taking the maximum and minimum correlation times as $f_H=1$ and $f_H=0$. In the case of the hydrodynamic radii (DLS) data this assignment is reversed. When applying Eq.~(\ref{eq:fraction}), with $G_H=g_HM_H$ and $M=M_H$, we find fitted values of $M$ for the correlation and scattering data to be 1.2 and 2.9 respectively. See Fig.~\ref{Coil_Glob}a. This is in stark contrast to the number of repeating groups that compose the polymer, around 1000, and with the results obtained from micellization in the previous section, which points to values of $M$ of the same order of magnitude as the number of ionizable groups. It is worth noting, just as the authors have too, that there is not enough data in the transition region for the scattering data to be confidently fit however. 
In Fig.~\ref{Coil_Glob}b the similarly scaled dynamic light scattering and pyrene probe fluorescent measurements from \cite{Eum1989} have been plotted. The maximum and minimal values of the fluorescent intensity are taken as $f_H=1$ and $f_H=0$, while this assignment is again reversed for the hydrodynamic radius data. Both sets of data present a very similar transition $pH$ and similar steepness in the curves. The fitted values of $M$ are 3 and 6 for the fluorescence and scattering data respectively, which is again considerably lower in order of magnitude than the number of carboxylic acid groups per chain, which is around 360.
As expected, the hydrophobic energy per group is lower for the PMAA than the PEAA, which is reflected in the lower transition $pH$. In both cases there is a very significant difference (2-3 orders of magnitude) between the effective value of $M$ and the number of repeating groups on the chains. As mentioned in the last section, the value of $M$ does not mirror the number of repeating units but rather the ionization change during the transition. \\
When comparing the $\theta$ and $1-f_H$ curves it is clear that the two quantities are not that well correlated. As the $pH$ is reduced, a significant amount of the polymer becomes deprotonated before the transitions takes place. This reduces the effect of the ionization change on the steepness of the transition. Moreover, a spreading out of the coil to globule transition with respect to $pH$ might be expected due to the assumption that there is a clear cut conformation change. As convincingly shown in \cite{Ulrich2005} a cascade of different conformations are seen (by computer simulation), such as so-called pearl-necklace conformations with local ‘pearls’ of collapsed states that are connected via ionized strings of the polymers \cite{Dobrynin1999, Kiriy2002, Blanco2019}. Several of these states are present between the limiting coil and globule conformations \cite{Ulrich2005}. See \hyperref[SI_broadening]{\emph{SI: Influence of intermediate states in transition broadening}} for further discussion. For single chains (and the associated globules) there will be a significant proportion of surface groups with respect to groups in the bulk of the aggregate. This allows for some of the ionic groups to remain ionized even when in the globule state, reducing the value of $M$. Although there are indications, as has been noted extensively in the literature, that the coil to globule transitions of hydrophobic polyelectrolytes are to a certain extent cooperative they do not seem to warrant a two-state treatment. A progressive transition including several intermediate states seems more likely. From these results we conclude that the coil-globule (-like) transition in HPE has the rudimentary signature of a cooperative transition but it is not in agreement with the MWC mechanism based on two conformation states where strong(er) coupling between conformation states ($f_H$) and ligand binding (ionization state $\theta$) is predicted. In contrast, in the situation with micelles formed by diblocks, the HPE blocks in the micelle cores correspond to a well-defined conformation that is clearly distinguishable from the aqueous coil state in the diblock system. We conclude from this comparison that intrinsic or other conditions are necessary to select well-defined conformation states. Such intrinsic condition can be 'programmed' in the architecture of the HPE, in this case in the form of HPE being linked to hydrophilic blocks  so that aggregation in the form of micelles is preferred over several intermediate states such as those in the coil-globule transition as illustrated in \cite{Ulrich2005}. A micelle interior can be seen as a reservoir that stabilizes the hydrophobic conformational state of a HPE. In principle, hydrophobic reservoirs can also be provided by other species that may stabilize one or more conformation states of the HPE (see Fig.~\ref{fig:partitioning}). In the next section we will show that the presence of lipid bilayers, in the form of (single or multilamellar) vesicles, can provide conditions in the form of reservoirs to select well-defined conformation states of HPE.\\

\subsection*{Membrane solubilization by disk formation}\label{Membrane_section}
\begin{figure}[h]
    \centering
    \subfloat[\centering Fitted $f_{aq}$ curves (Eq.~(\ref{diskfraction})). Global parameters: $M=65$ , $\Delta g_{EEA}=0.49k_BT$, $\Delta g_{MAA}=0.22k_BT$ and ${\frac{M_{D}}{M}}=0.9$. $pK_a$= 4.5 is assumed. Inset: transition $pH$ trend with respect to the EAA mole fraction of the polymer ($x$). Fitted using Eq.~(\ref{pHda}).]{{\includegraphics[width=0.5\linewidth]{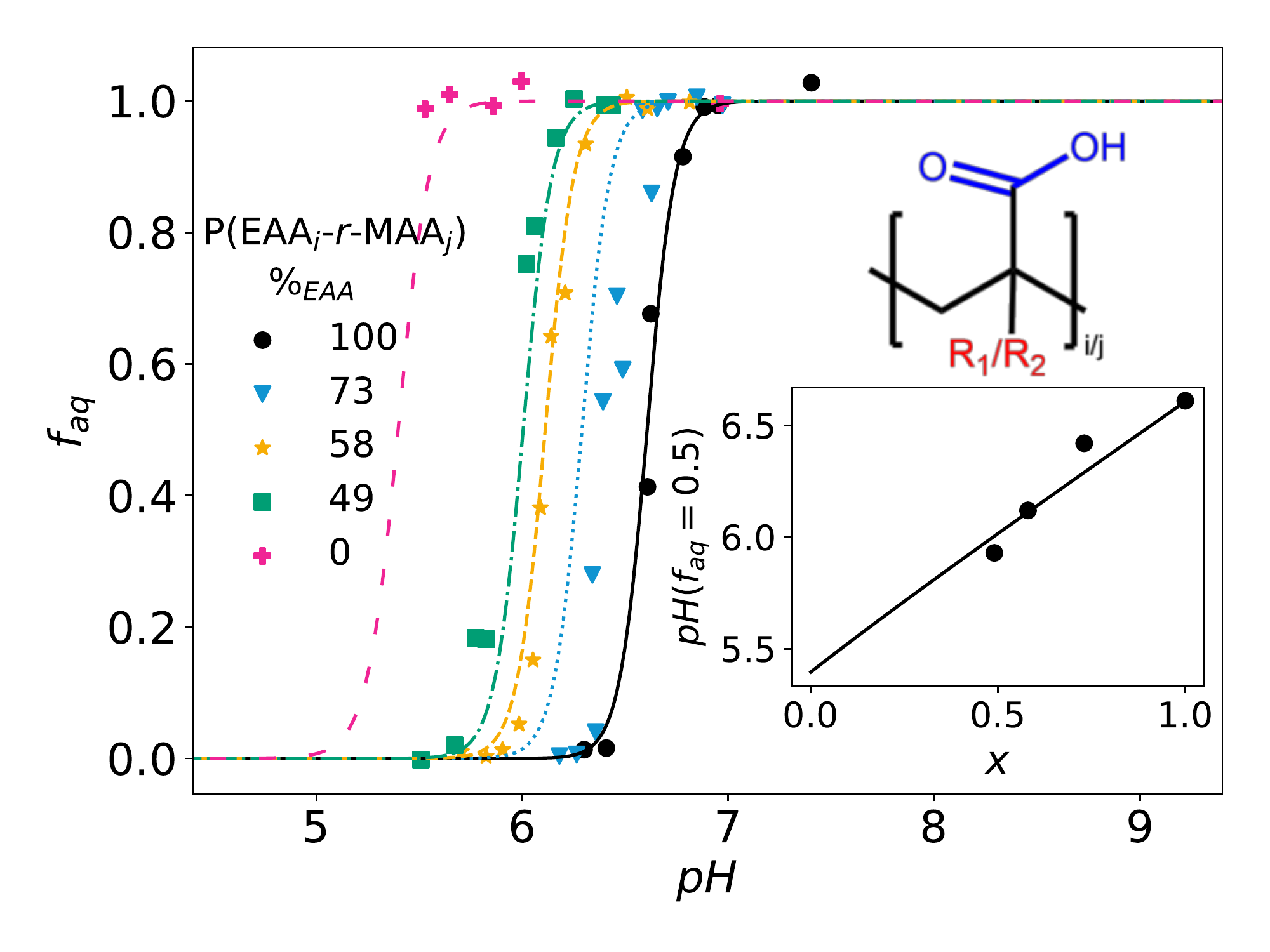} }}
   \subfloat[\centering Fitted $f_{H}$ curves (Eq.~(\ref{HDfraction})). Local parameters: $M=19.5, 14.8, 20$ for the ${\frac{M_H}{M}} =$~2, 3 and 4 variants respectively. Global parameters: $g_H=1.28k_BT$ and $pK_a= 4.3$. Note $M_H/M=j/i$. Inset: transition $pH$ fitted with  Eq.~(\ref{pHhd}).]{{\includegraphics[width=0.5\linewidth]{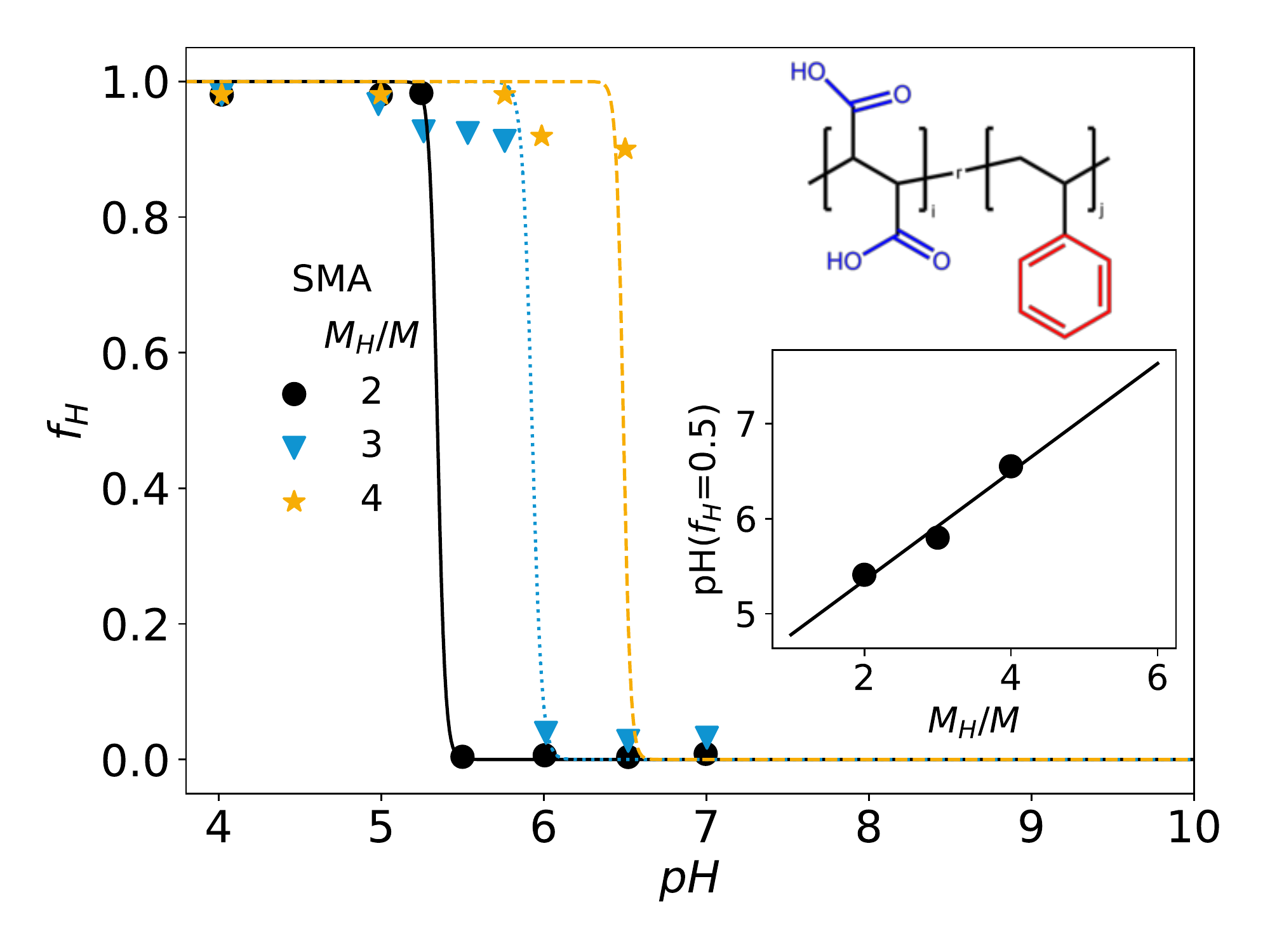} }}%
    \caption{Membrane dissolution data for (a) PEAA-PMAA \cite{Thomas1995} and (b) SMA \cite{Scheidelaar2016}. The chemical structures of the polymers are given as insets.}
	\label{fig:Disks}
\end{figure}

\begin{figure*}[t]
    \centering
    \includegraphics[width=\linewidth]{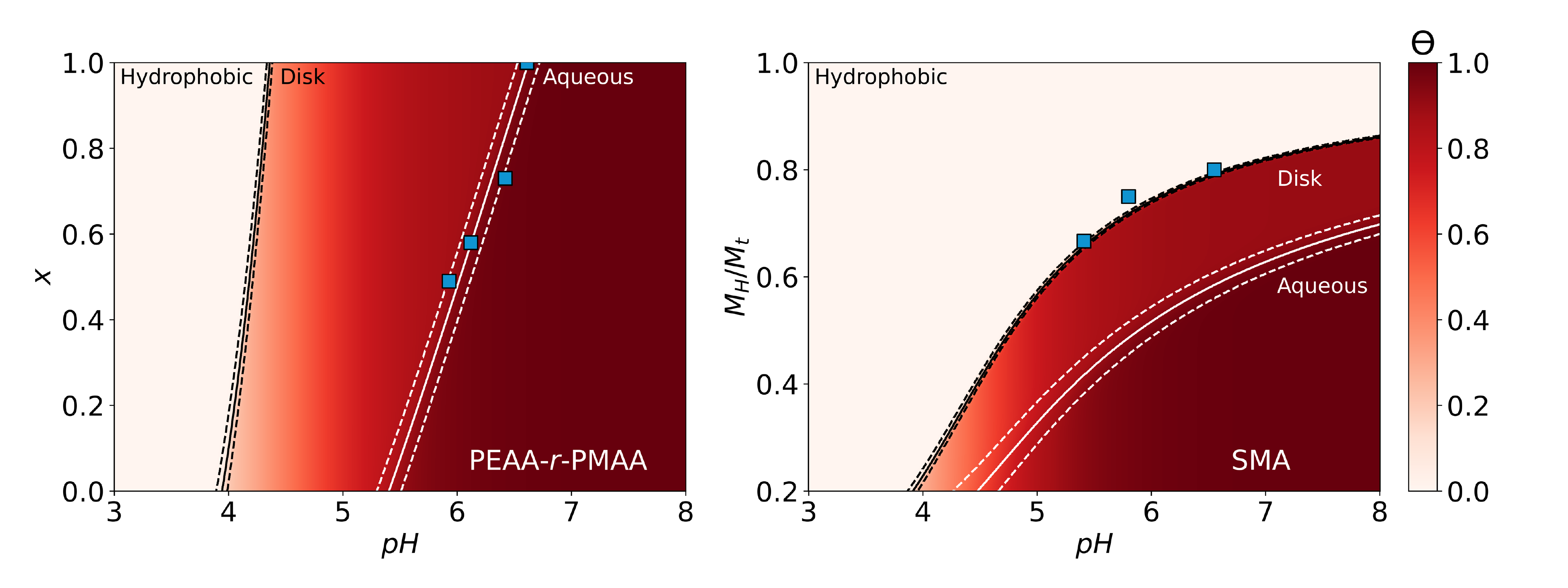}
    \caption{Calculated phase diagrams for HPE as a function of $pH$ and composition for the PEAA-r-PMAA system (left) and the SMA system (right) using Eq.~(\ref{diskfraction},\ref{HDfraction}). Solid lines (black for the hydrophobic-disk transition and white for the disk-aqueous transition) denote the transition between the different phases and dotted lines where the main phase reaches a fraction of 0.9. Note the marked difference in transition steepness for the disk-aqueous versus the hydrophobic-disk transition. The blue squares are experimental data at the transition midpoints in Fig.~\ref{fig:Disks}. The red hue in the background of the figures reflects the ionization fraction, $\theta$ (Eq.~(\ref{thetaD})), of the HPE. Global parameters: Left: $g_{MAA} = \text{0.44}k_BT$, $g_{EAA} = \text{0.98} k_BT$, $G_{HD} = G_H/2$ , $pK_a=4.5$ and $M=100$. $x$ is the mole fraction of EAA in the polymer. Right: $g_H=\text{1.6}k_BT$, $g_{HD}=g_H/\text{1.3}$, $pK_a=4.3$ and $M_H+M=M_t=100$.}
	\label{fig:Diagrams}
\end{figure*}

Many hydrophobic polyelectrolytes are known to interact with lipid bilayer membranes \cite{Yessine2004}. Depending on the hydrophobicity, chemical structure and relative concentration of the HPE, membranes exhibit solubilization or fusion among other destabilization mechanisms. These processes, usually triggered via a $pH$ change, tend to be coupled to the release of the contents of the membrane, leading to the interest in such systems for drug delivery applications. Styrene-maleic acid (SMA) is used to make nanometer sized ($\sim10$ nm) vesicle nanodisks which allow for the study of membrane proteins in their local environments \cite{TONGE,knowles,Dorr2016}. The bottom row of Fig.~\ref{fig:partitioning} shows a schematic of a disk and the states of an HPE in such a system. Tirrell and coworkers showed in a series of papers the solubilization of different types of lipid vesicles by poly(ethylacrylic acid)-r-poly(methylacrylic acid) copolymers (PEAA-r-PMAA)\cite{Thomas1995}. In Figure \ref{fig:Disks} we summarize the findings in \cite{Thomas1995, Scheidelaar2016} for these different HPE.
\\The stabilization of a nanodisk phase can be seen as analogous to the stabilization of any hydrophobic-hydrophilic interface, albeit one with a high curvature. Therefore we can expect the hydrophobic polyelectrolyte to act like a surfactant with the hydrophobic groups directed towards the core of the nanodisk and the carboxyl groups pointing out into solution. This mechanism relies on the two moieties being able to freely rotate with respect to the chain backbone, as in styrene-maleic acid copolymers, or be located on different branches of the same monomer. The family of poly(alkyl acrylic acid) polymers fall under that category. See Table~S1 for a comparison of some HPE structures.\\
Tirrell et al. describe the $pH$-dependent solubilization of DPPC multilamellar membranes using PEAA-r-PMAA copolymers with a chain length of around 2000 for all of the polymer compositions. They focus on the tunability of the transition $pH$ on changing the EAA to MAA ratio in the polymer. 
Turbidimetry was used to monitor the dissolution of the membranes as function of $pH$. The measurements are normalized to the value measured before the solubilization transition where unperturbed membranes are present. Therefore, assuming no intermediate structural changes other than disk formation, the turbidity measured directly correlates with the fraction of membranes that remain undissolved. Only the aqueous to disk transition was investigated therefore only the difference in hydrophobic penalty between the solubilized polymer and the polymer at the disk solvent interface is needed to characterise the system.
In absence of the usual reference for the hydrophobic penalty, that is the penalty in the hydrophobic state, only a difference in hydrophobic penalty between the aqueous and disk state can be extracted. This is analogous to making the disk state the reference state.\\
The $pH$ at which the transition from the disk to the aqueous state takes place, $pH_{DA}$, follows from setting $\Xi_{aq} = \Xi_D$ in Eqs.~(\ref{ksiaq}, \ref{ksid}) which leads to $pH_{DA} = pK_a + \frac{\text{0.4343}\beta (G_H - G_{HD})}{(M - M_D)}$. We write the hydrophobic free energy difference between the aqueous state and the disk conformation as a function of composition via $G_H - G_{HD} = M_H f(\Delta g_{EAA}, \Delta g_{MAA}, x)$. Here $f(\Delta g_{EAA}, \Delta g_{MAA}, x) = x \Delta g_{EAA} + (1-x) \Delta g_{MAA}$, where  $\Delta g_{EAA}, \Delta g_{MAA}$ are the hydrophobic free energy differences between the aqueous state and disk conformation per monomer, and $x$ is the mole fraction of EAA in the polymer. Combining all that leads to (note that we have $M_H = M$ here)
\begin{linenomath}
\begin{equation}
pH_{DA} = pK_a + \frac{\text{0.4343} \beta f(\Delta g_{EAA}, \Delta g_{MAA}, x)}{1 - \frac{M_{D}}{M}} \label{pHda}.
\end{equation}
\end{linenomath}
The midpoints of the transition are found using a trial fit of the data where $G_{H} - G_{HD}$ and $M - M_D$ are free parameters, see Eq.~(\ref{diskfraction}) below. A value of 4.5 is assumed for the $pK_a$. We fixed $M_D/M =$~0.9, which is purely an assumption: 10\% uncharged groups in the disk conformation compared to the aqueous state seems a reasonable upper limit. With that, a value for $\Delta g_{EEA}$ is trivially found using the midpoint value for $x$=1 and linear regression can be used to find the value of $\Delta g_{MAA}$ from Fig.~\ref{fig:Disks}a.
Inserting the derived hydrophobic penalty values and  the assumed value of the $pK_a$ we calculate the disk (and aqueous) fraction by
\begin{linenomath}
\begin{multline}
	f_D =1 - f_{aq} = \Xi_D/(\Xi_{aq} + \Xi_D)\\
	= \left(1 + \text{exp}[\beta (G_{HD} - G_H)](1 + 10^{(pH - pK_a)})^{(M - M_D)} \right)^{-1}.  \label{diskfraction}
\end{multline}
\end{linenomath}
In calculating the fractions we used an average effective value of $M - M_D$ = 6.5. The value of $M$ cannot be extracted independently at this point. 
The match between the fitted transition and the experimental data is reasonable although not as good as in the micelles in Fig.~\ref{comp_fig}. This is because there are stronger deviations from the linear relation between transition $pH$ and $x$, as can be seen in the inset in Fig.~\ref{fig:Disks}a.\\
Scheidelaar et al. report the solubilization of DMPC by styrene-maleic acid random copolymers of different compositions \cite{Scheidelaar2016}. The disk to interbilayer spacing (hydrophobic) transition is remarkably sharp, considering that the average length of the polymers is of a few tens of units. The data clearly demonstrates how oligomeric species are also capable of extremely sharp $pH$-induced transitions.
We find the transition $pH$ for polymers with a ratio of styrene over maleic acid $M_H/M$ by using Eq.~(\ref{ksid}) and setting $\Xi_D = \Xi_H = 1$. Taking $G_{HD} = M_H g_{HD}$ leads to
\begin{linenomath}
\begin{equation}
	pH_{HD} = pK_a + \text{0.4343}\beta g_{HD} \frac{M_H}{M}  \label{pHhd}.
\end{equation}
\end{linenomath}
The fraction of HPE in the hydrophobic conformation, ignoring the weight of the aqueous conformation around this transition, is given by
\begin{linenomath}
\begin{align}
	f_H =\frac{\Xi_H}{\Xi_H + \Xi_D} &= \left(1 + \text{exp}(-\beta G_{HD})(1 + 10^{(pH - pK_a)})^{M_D} \right)^{-1} \nonumber \\
	&\text{and}~f_{D} = 1 - f_H.  \label{HDfraction}
\end{align}
\end{linenomath}
This analysis (Fig.~\ref{fig:Disks}b) follows the expected trend and yields a value for $g_{HD}$ of 1.28$k_BT$ and an effective value of the $pK_a$ of 4.3. The value for the effective $pK_a$ is in good agreement with the $pK_a$ of the first ionization of succininc acid, 4.21 \cite{Lide2005} (note that maleic acid in the polymerized form in p(SMA) is structurally closer to succinic acid than to maleic acid). The second ionization of succinic acid has a higher $pK_a$ of 5.6 and therefore might also play a role in these transitions, especially for the more hydrophobic polymers with higher values for the transition $pH$. In the SI, we report a similar analysis of the (macroscopic) aggregation transition in SMA, also from Ref. \cite{Scheidelaar2016}. There an effective $pK_a$ of 2.4 is found. However as can be seen in Fig.~S3 the ionization and the aggregation state of the polymer are largely uncorrelated. A calculation of the effective $pK_a$ for this system will not yield meaningful values with respect to the ionization state of the polymer. The analysis in the \hyperref[SI_aggregation]{\emph{SI:HPE Aggregation}} also reveals a value for the hydrophobic penalty relative to the aggregated state: $\beta g_H \approx 2.1$ which is significantly higher than the value of $\beta g_{HD} \approx 1.3$ that we find. A lower hydrophobic penalty $g_{HD}$ is indeed expected due to the postulated penetration of the hydrophobic groups into the rims of the nanodisks. Note that this value also includes the work of formation of disks out of bilayer membranes.\\  
The effective values of $M_D$ for the 2:1 and 3:1 variants were 19.5 and 14.8 respectively. Due to the overlap with what seems to be the aqueous to disk transition data, no attempt at estimating the sharpness was carried out for the 4:1 variant. The effective values of $M_D$ point to a transition that is more cooperative (sharper) than the aqueous-disk transition in the PEAA-r-PMAA copolymers, despite the much longer chain length of the latter. While we cannot rule out other broadening effects, this effect is at least partly due to the fact that the sharpness of the aqueous-disk transition is governed by the difference $M - M_D$ (Eq.~(\ref{diskfraction})) while sharpness of the hydrophobic disk transition (as in SMA) is measured by the value of $M_D$ (Eq.~(\ref{HDfraction})). Finally, in comparison with the aggregation transition of SMA analyzed in the SI, where we find values of $M$ of 5.6, 4.9, and 12.0 for the 4:1, 3:1 and 2:1 styrene-maleic acid ratios respectively, the hydrophobic-disk transition is significantly sharper. That again points to the requirement of well-defined reservoirs (in the form of bilayers) that are able to stabilize a finite number of conformational states (here the disk and presumably the hydrophobic state where the HPE are dissolved in the interbilayer spacings). Comparison between the aqueous fractions and ionization states in the aggregation transition indeed reveals a similar lack of correlation as the situation for the globule-coil transition.\\
In Fig.~\ref{fig:Diagrams} the results in Fig.~\ref{fig:Disks} are summarized and complemented with predicted scenarios for the transition from hydrophobic to disk in Fig.~\ref{fig:Diagrams} (left), and from disk to aqueous in Fig.~\ref{fig:Diagrams} (right). These predicted phase diagrams are based on (at this point) unverifiable choices for the hydrophobic contributions. We also show the predicted behavior of the ionized fraction Eq.~(\ref{thetaD}), again based on the assumption that 10\% of the ionizable groups remain uncharged in the disk conformations.
The experiments in Ref.~\cite{Scheidelaar2016} indicate that the aqueous to disk transition presents a much more noisy, and much less sharp and well-defined transition compared to the hydrophobic-disk transition. This might be expected due to the ionization difference between the aqueous state and the disk state potentially being small (10\%), and is indeed confirmed in the theoretical 'phase diagram' in Fig.~\ref{fig:Diagrams} (right). The experiments in Fig.~6b in \cite{Scheidelaar2016} indeed seem to point to a broad transition at $pH$ values above approximately 7. The prediction is that longer chain length of SMA will lead to a sharper aqueous-disk transition. In order to pin down the values of the hydrophobic free energies and the uncharged fraction of ionic groups, the ionization state of the HPE around the transitions needs to be known. The experimental determination of that quantity is expected to be challenging as the ionization state of the head groups of the lipids that make up the bilayers is also expected to (slightly) depend on $pH$.\\
In closing this section we would like to mention that the MWC-like mechanism in this work has not been included in the term 'cooperative' as discussed in \cite{Ercolani2003}. There, the term 'cooperative' has been reserved for situations where binding sites interact. We are aware that our used terminology can be debatable. As we see it, the transitions we describe in this work occur, and only occur, because of the binding of several ligands at once, and the term 'cooperative' therefore seems appropriate. We note that the influence of Coulomb interactions in weakening the transitions is a form of 'negative cooperativity' due to interactions between bound ligands. Both types of cooperativity add up and contribute to the effective value of the degree of cooperativity $M$ in the several scenarios that we investigated.
\section*{Conclusion}
The examples laid out in the previous sections illustrate that the transitions carried out by hydrophobic polyelectrolytes can range from strongly to weakly cooperative. Analysis of the micellization transition in diblocks provide strong evidence that the underlying mechanism of the observed cooperativity is in agreement with the MWC model \cite{Monod1965} that was originally designed to understand allosteric transitions. We verify here that the MWC model is more general: allostery, or interactions between binding sites, is not a requirement for the MWC model to work. What is required is the availability of two or more well-defined conformations with different affinity for ligands (here protons or hydroxyl ions). In the relatively simple substrates (at least compared to hemoglobin) studied here, the conformational states are coupled to hydrophobic or aqueous reservoirs. These reservoirs may be self-induced, such as in the case of micelles, or due to external structures being present, such as during the solubilization of bilayers. In the HPE, the conformational penalty in the aqueous and disk state is (within reasonable accuracy) a linear combination of the composition of the polymers. This can be clearly seen in Fig.~\ref{comp_fig} for the micellar systems and Fig.~\ref{fig:Disks} for the disk formation. While this points to MWC as a plausible mechanism for disk formation, more quantitative comparison between theory and experiments is desired. Additional experiments to make that possible are for example the determination of the ionization states of HPE in the disk and aqueous states, as shown in the form of predictions in Fig.~\ref{fig:Diagrams}, as well as the typical adsorption density of HPE onto the disk rims. Moreover, for medical applications it would be relevant to study the influence of temperature. \\
In principle the observed cooperativity, as well as the ability to tune the transition $pH$ in disk formation may provide a strategy to specifically target tumor cells, see the \hyperref[SI_applications]{\emph{SI:Applications}} for a discussion of this possibility. There likely are several hurdles to overcome, for example dilution effects and the unknown role of membrane proteins. The principles laid out in this work are not only applicable in 'simple' HPE but may also be applied in designing oligo peptides with combined hydrophobic and acidic (or zwitterionic) amino acids, see, e.g.,\cite{Anufrieva1968b, Anufrieva1975, Stokrova1989}. These types of oligopeptides, often referred to as 'cell penetrating peptides' \cite{Copolovici2014}, depending on their architecture, may permeate cell membranes as a function of $pH$ or by the concentration of ligands other than protons. \\
Weakly crosslinked HPE have also been observed to have a sharp $pH$-mediated transition: from a swollen (with water) to a collapsed state, see, e.g., Ref. \cite{Siegel1988, Siegel1993}. We expect that there, the swollen state is analogous to the 'aqueous' conformation, and the collapsed state is similar to the 'hydrophobic' state in the previous sections. By being crosslinked, the occurrence of many intermediate conformations between the aqueous and hydrophobic states may be avoided. Potential applications of these systems are for example actuator, optical switches and drug delivery vehicles that are driven by small $pH$ variations. It should be noted that details can be important here as not all crosslinked HPE have a narrow transition, see, e.g., \cite{Philippova1997}. There, crosslinked HPE are being studied that consist of relatively strongly hydrophobic alkyl-acrylates with alkyl chain lengths between 8 and 18 carbon units, which may lead to local phase separation withing the gels.\\
In general it is expected that (macro) molecular substrates that have multiple binding sites for ligands can undergo sharp transitions driven by small variations in ligand concentration. The degree of sharpness, or cooperativity, is largely determined by the stabilization of well-defined conformational states. Besides the acid-base systems analyzed in this work, we expect that other relatively simple host-guest systems \cite{lehn_1995}, can have similarly sharp transitions. In these systems, the 'hosts' are functional groups on (hydrophobic) oligomers or polymers, and the 'guests' are ligands.

\section*{Acknowledgements}{We thank Jinming Gao and Yang Li for kindly providing the data on HPE diblocks, and Neshat Moslehi, Bas van Ravensteijn and Tina Vermonden for discussions on hydrophobic polyelectrolytes. Antoinette Killian, Adrian Kopf, and Martijn Koorengevel are thanked for several illuminating brainstorm sessions on membrane disk formation. WKK thanks Rob Phillips and Tal Einav for enlightenment regarding MWC theory. Finally we acknowledge the Dutch Research Counsil (NWO) for funding (grant no 712.018.003).}

\section*{Supporting information}
\setcounter{equation}{0}
\setcounter{figure}{0}
\renewcommand{\thefigure}{S\arabic{figure}}
\renewcommand{\theequation}{S\arabic{equation}}
\renewcommand{\thetable}{S\arabic{table}}
\subsection*{Theory}\label{SI_theory}
\subsubsection*{Generalization of the Monod - Wyman - Changeux (MWC) model}
\begin{figure}[h]
	\centering
	\includegraphics*[width=15cm]{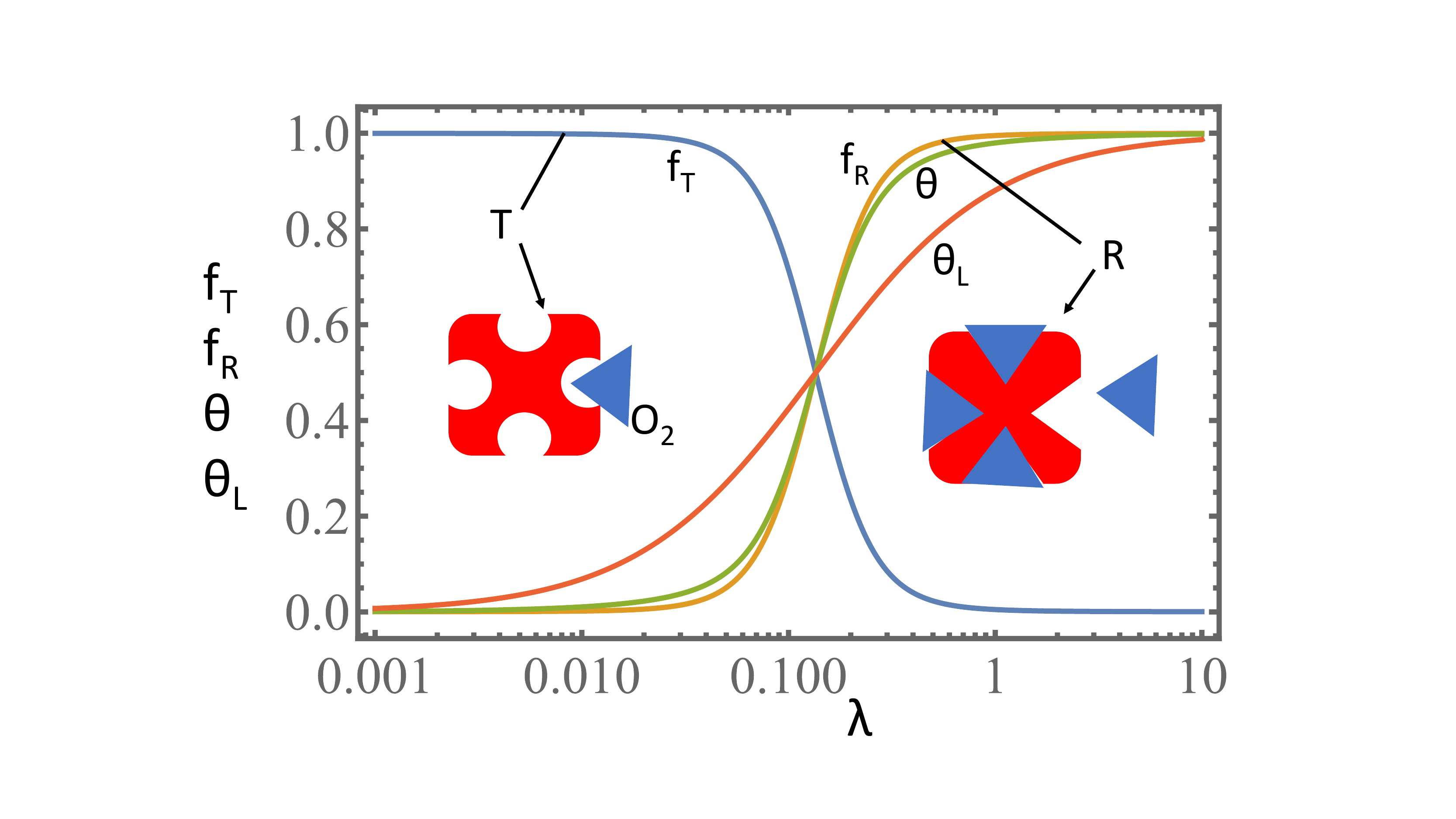}
	\caption{Plot of the fraction of hemoglobin in the Relaxed ($f_R$) and Tense ($f_T$) state, Eqs.~(\ref{fp}), and the fraction of bound oxygen onto hemoglobin, Eq.~(\ref{SI_theta}), as a function of $\lambda$. $\lambda$ is proportional to the partial oxygen pressure. In calculating the fractions $f_T, f_R$ and the fraction of oxygen bound to the $M=4$ binding sites, $\theta$, we used in Eqs.~(\ref{ksip}, \ref{SI_theta}) $G_T^* = g_T =0$, $\beta G_R^* = 8$, $\beta g_R = -4$. The Langmuir isotherm $\theta_L$ is Eq.~(\ref{Langmuir}) with $\beta g_1 = -6$.}
	\label{fig:Hb}
\end{figure}

The MWC model \cite{Monod1965, Monod1963} addresses allosteric transitions by considering two different conformational states of a (macro) molecule, or substrate, that is able to bind to one or several ligand molecules. Each conformational state of a substrate has a different binding strength of the binding sites with the ligands. In the MWC mechanism, the ground state of the substrate has a relatively weak binding affinity for a ligand. The other conformation of the substrate is unfavorable in the absence of ligands, but has a relatively strong binding affinity for the ligands. This mechanism can lead to a sharp transition in ligand occupancy of the substrate as a function of ligand concentration or partial pressure of the ligand. A classical example is the binding of oxygen onto hemoglobin. A hemoglobin molecule can be in two conformational states. One is the tense (T) state being the conformational ground state, which has a relatively weak affinity for oxygen. The relaxed (R) state is conformationally unfavorable but has a relatively strong affinity for oxygen. See the insets  in Fig.~\ref{fig:Hb} for a schematic drawing of the situation. The competition between the two states leads to a sharp increase in the number of bound oxygen molecules per hemoglobin molecule as a function of oxygen pressure. However, there are many more situations in biology where the MWC model applies, for example, ligand-gated ion channels in cell membranes, genome accessibility of transcription factors, and bacterial chemotaxis, see \cite{Marzen2013} for a review. In this section we write the MWC model in the language of statistical mechanics and generalize the model to an arbitrary number of conformational states of a substrate. The substrate can be a protein or another type of macromolecule. For the sake of generality we (for now) do not specify the nature of the ligands - these are molecules that bind to the binding sites of the substrates. Each conformational state of a substrate has its characteristic ligand affinity and conformational penalty. Subsequently we apply the result to hydrophobic polyelectrolytes, where protons or hydroxyl ions act as ligands.\\ 
We write the coarse-grained grand partition function of a substrate that can be in $P$ different conformational states as
\begin{equation}
	\Xi =\sum_{p=1}^{P}\sum_{N=0}^{M}\lambda^N Z_p(N,T,M)  \label{ksi}
\end{equation}
In this equation, $\lambda = \exp{(\beta \mu)}$ with $\beta = 1 /k_B T$. $k_B$ is Boltzmann's constant and $T$ the absolute temperature. $\mu$ is the chemical potential of the ligand that adsorbs (or binds) onto the substrate. $\mu$ is related to the ligand concentration or partial pressure of the ligand. The coarse-grained canonical partition function of a substrate with $M$ binding sites in its conformational state $p$ and with $N$ bound ligands is given by
\begin{equation}
Z_p(N, T, M) =\exp{(-\beta G_p^*)} {M \choose {N}} \exp (- \beta N g_p) \label{Z}
\end{equation}
Here, $G_p^*$ is the conformational penalty of the substrate in state $p$. $g_p$ is the binding free energy, or affinity, of a ligand onto the substrate in state $p$. In writing down Eq.~(\ref{Z}) we assumed uncorrelated binding of the $N$ ligands, that can bind with a multiplicity ${M \choose {N}} \equiv M!/N!(M-N)!$ onto the $M$ binding sites of a substrate. Using that we apply the binomial theorem and find
\begin{equation}
	\Xi =\sum_{p=1}^{P} \exp{(-\beta G_p^*)} \left(1 + \lambda \exp{(-\beta g_p)} \right)^M = \sum_{p=1}^P \Xi_p.  \label{ksi2}
\end{equation}
Here we defined the grand partition function of a substrate in its conformational state $p$ as
\begin{equation}
	\Xi_p = \sum_{N=0}^{M}\lambda^N Z_p(N,T,M) = \exp{(-\beta G_p^*)} \left(1 + \lambda \exp{(-\beta g_p)} \right)^M.  \label{ksip}
\end{equation}
Eq.~(\ref{ksip}) is the statistical weigth of a substrate in conformation state $p$. We obtain the fraction of substrates in the $p$ state, $f_p$, by
\begin{equation}
	f_p=\frac{\Xi_p}{\Xi}			\label{fp}
\end{equation}
The average fraction of bound ligands on a substrate is given by
\begin{multline}
	\theta = \frac{\langle N \rangle}{M} = \frac{1}{M} \frac{\sum_{p=1}^{P}\sum_{N=0}^{M}N \lambda^N Z_p(N,T,M)}{\Xi}= \frac{1}{M} \frac{\lambda}{\Xi} \frac{\partial \Xi}{\partial \lambda} = \\
	= \lambda \Xi^{-1} \sum_{p = 1}^P \exp{(- \beta (G_p^* + g_p))} (1 + \lambda \exp{(- \beta g_p) })^{(M-1)} \label{SI_theta}
\end{multline}
By setting $P=1$, that is, with only a single conformational state available for the substrate, Eq.~(\ref{SI_theta}) reduces to the Langmuir adsorption equation, as it should:
\begin{equation}
 \theta_L = \frac{\lambda \exp{(-\beta g_1)}}{1 + \lambda \exp{(-\beta g_1)} }.~~~~~\text{Langmuir adsorption equation} \label{Langmuir}
\end{equation}
The situation for oxygen binding onto hemoglobin follows by setting $P=2$, $M=4$, with the ground state ($p=1$) labeled as the Tense (T) state with binding affinity $g_T$ and conformational penalty $G_T^* = 0$. 
The Relaxed (R) state ($p=2$) has binding affinity $g_R$ and conformational penalty $G_R^* > 0$. In Fig.~\ref{fig:Hb} we apply the model to the situation for oxygen binding onto hemoglobin. $\lambda$ is proportional to the partial oxygen pressure. The red objects in the inset are schematic drawings of the hemoglobin molecule. In the tense (T) state, hemoglobin is in its ground state, being the case at low values of $\lambda$. There, affinity for oxygen (blue triangle) is low which is illustrated by the geometry of the binding sites onto which oxygen does not fit well. At relatively high oxygen pressure (larger values of $\lambda$), the Relaxed (R) state, with relatively strong affinity for oxygen binding (the shape of the binding sites now provide a good fit with oxygen), becomes stable. There is a sharp crossover from the fraction of hemoglobin in its tense state, $f_T$, to its relaxed state, $f_R$, which correlates strongly with the fraction of bound oxygen to hemoglobin ($\theta$). As a comparison we plotted the situation for a single available conformation, $\theta_L$ (Eq.~(\ref{Langmuir})). Clearly the crossover from $\theta_L = 0$ to $\theta_L = 1$ as a function of $\lambda$ is much more gradual (or less steep) as compared to the MWC mechanism for hemoglobin. This example illustrates that the MWC model can explain the steep transitions in binding fraction as a function of ligand concentration. The transition is accompanied by a comparibly steep crossover from the ground conformational state of the substrate to the state with stronger binding affinity. These transitions can be seen as cooperative, that is, the transition occurs upon binding of multiple ligands onto the substrate. Interestingly, interactions between binding sites and/or bound ligands are not required in order to explain this class of cooperative transitions.
\\
As described in the main text, hydrophobic polyelectrolytes (HPE) consist of hydrophobic as well as ionizable (acidic or basic) functional groups. We consider these polymers in their ground state when they are in a hydrophobic ('oily') reservoir with their ionizable groups being (mostly) uncharged. The hydrophobic reservoir can be the interior of a micelle in water in the situation where the HPE are linked to hydrophilic polymers. It can also be the interbilayer spacing in lipid vesicles (see Fig.~\ref{fig:partitioning} in the main text), or, in principle, a separate, macroscopic, 'oil' phase that is in contact with a macroscopic aqueous phase. The main difference from the situation with hemoglobin and other proteins is that the conformational state of a HPE now couples to a reservoir rather than to some intrinsic folding transition that influences the binding affinity for ligands. With that in mind we apply the model for MWC transitions to HPE. The hydrophobic ground state has no conformational penalty so that for that state, with effectively $M$ ionizable groups on a HPE, Eq.~(\ref{ksip}) reads $\Xi_H = (1 + \lambda \exp{(- \beta g_h)})^M$. Here $\lambda$ is a measure for the concentration of hydroxyl ions or protons, and $g_h$ is the binding affinity of acidic or basic functional groups for the ions in the hydrophobic state. We assume that in this 'oily' state, $\beta g_h >> 1$, that is, ionization in hydrophobic media is unfavorable, and thus, for a HPE in its hydrophobic state or reservoir,
\begin{equation}
	\Xi_H = 1.		\label{ksiH}
\end{equation}
Another conformational state is when HPE are dissolved in aqueous solution. This state is expected to become favorable upon charging of the acidic or basic functional groups of these molecules as binding sites for, respectively, hydroxyl ions or protons. The hydrophobic groups of these polymers are responsible for the conformational penalty. We write the statistical weight of the aqueous state, via Eq.~(\ref{ksip}) as $\Xi_{aq} = \exp{(- \beta G_H)} (1 + \lambda \exp{(-\beta g)})^M$. Here $G_H \equiv G_{aq}^*$ being the conformational penalty of a HPE in the aqueous state, caused by the hydrophobic character of the hydrophobic functional groups of the HPE. Further $g \equiv g_{aq}$ stands for the binding free energy of an ionizable group for hydroxyl ions or protons. For carboxyl groups with hydroxyl 'ligands' we write $\lambda \text{exp} (- \beta g) = 10^{pH - pK_a}$, with $pK_a = - ^{10}\text{log}K_a$, $K_a$ being the dissociation constant of a (solvated) carboxyl group. Using that we find 
\begin{equation}
	\Xi_{aq} = \text{exp}(-\beta G_H)~(1 + 10^{pH - pK_a})^M.~~~\text{(carboxyl ionizable groups)} \label{ksiaqac}
\end{equation}
Similarly we find for basic groups 
\begin{equation}
	\Xi_{aq} = \text{exp}(-\beta G_H)~(1 + 10^{pK_a' - pH})^M.~~~\text{(basic ionizable groups)} \label{ksiaqbs}
\end{equation}
In this equation, $pK_a' = - ^{10}\text{log}K_a'$, $K_a'$ being the dissociation constant for the conjugate acid of the basic group.\\
\subsubsection*{HPE bound to lipid bilayer disks} 
We define a third conformational state of the polymer when adsorbed onto disks, see the main text \hyperref[theory]{\emph{Theory}} section. In the disk state, we assume that the hydrophobic parts of the polymer still pay a penalty for being at a hydrophobic - aqueous interface but that penalty should be significantly smaller than for being fully in the aqueous state. Finally we take into account that the geometric constraints imposed on the polymer coupled to the thermal fluctuations of the chain, may lead to a fraction of the electrical charges on the ionizable groups being quenched when immersed in or very close to the hydrophobic bilayer region. Therefore, on average, less chargeable groups may get ionized compared to the situation where the polymers are fully dissolved in the aqueous state. The derivation of the grand partition function of the polymer in the disk state goes along the same line as for the aqueous state, Eq.~(\ref{ksiaqac}). Considering only carboxyl ionic groups, the result is     
\begin{equation}
	\Xi_{D} = \text{exp}(-\beta G_{HD})~(1 + 10^{pH - pK_a})^{M_{D}}, \label{SI_ksid}
\end{equation}
where $G_{HD}$ stands for the hydrophobic penalty of a HPE chain when adsorbed onto a disk which includes the formation free energy of the disk (per HPE chain) and $M_D$ is the number of chargeable groups on the HPE in the disk state. The value of $G_{HD}$ is expected to be a fraction of the value of $G_H$ in Eq.~(\ref{ksiaqac}). $(M - M_{D})/M$ should be seen as the fraction of time chargeable groups spend inside or close to the hydrophobic bilayer region.
\subsection*{Applications}\label{SI_applications}
\subsubsection*{Potential application of HPE in tumor treatment}
In this section we discuss a potential application of HPE in specifically targeting tumor cells. The extracellular $pH$ of healthy cells is 7.2, while that of fast-growing tumor cells is around 6.6-6.8. The situation is reversed for the intracellular pH: 6.8 in healthy cells and 7.2-7.4 in tumor cells. This is known as the Warburg effect \cite{Warburg1956, Sun2019}, which is caused by tumor cells often being in a fermentation-like metabolic mode. We discuss the possibility of tuning HPE in such a way that only tumor cells (with relatively low extracellular $pH$) are being solubilized or permeated while healthy cells remain unaffected, a strategy similar to pH responsive tumor-targeted drug delivery \cite{Kanamala2016}. An important question is whether cell membranes are similarly affected by HPE as a function of $pH$ as bilayer vesicles are. While plasma membranes of bacteria and eukaryotic cells consist of lipid bilayers, they also contain embedded membrane proteins that may interact with HPE. These interactions may lead to different behavior as compared to vesicle bilayers. Red blood cells have been observed to solubilize (hemolyze) upon addition of poly(propyl acrylic acid) (PPAA) at $pH <6$ and poly(ethyl acrylic acid) (PEAA) when $pH < 5$ \cite{Murthy1999}. This trend is consistent with our predictions in terms of the hydrophobicity (value of $g_H$ in Eq.~(\ref{pHda}) in the main text) of these polymers and with the observations in bilayer vesicles \cite{Thomas1992}, although it is not clear if disks are formed with the red blood cell membranes. Interestingly, with the most hydrophobic HPE (PPAA), the transition from no hemolysis to full hemolysis upon decreasing $pH$ occurs within a similar $pH$ range as in the bilayer vesicles in Ref. \cite{Thomas1992}, that is, a range of at most 0.5 $pH$ units \cite{Murthy1999}. This is promising with respect to the application we have in mind. However, there is a caveat in terms of generic behavior of cell membranes and HPA: a counter example is a study of E. coli membranes and yeast mitochondria with poly(styrene maleic acid) (SMA) in \cite{Kopf2020}. There, the behavior as a function of $pH$ seems qualitatively different from the observations with red blood cells and with bilayer vesicles. In particular, partial solubilization of E. coli cells and yeast mitochondria is being observed at high $pH$ ($>8$) and not (or much less) at lower $pH$. As discussed in \cite{Kopf2020}, positively charged membrane proteins may play a role, possibly in combination with the two ionizable groups onto the maleic acid residues of SMA. At this point it is not clear if the different behavior of SMA in combination with E. coli cells and yeast mitochondria, compared to the examples with eukayotic (red blood) cells and with lipid bilayer versicles, is caused by the special nature of E. coli and yeast mitochondria membranes, the properties of SMA, or both.
\begin{figure}[h]
	\centering
	\includegraphics*[width=15cm]{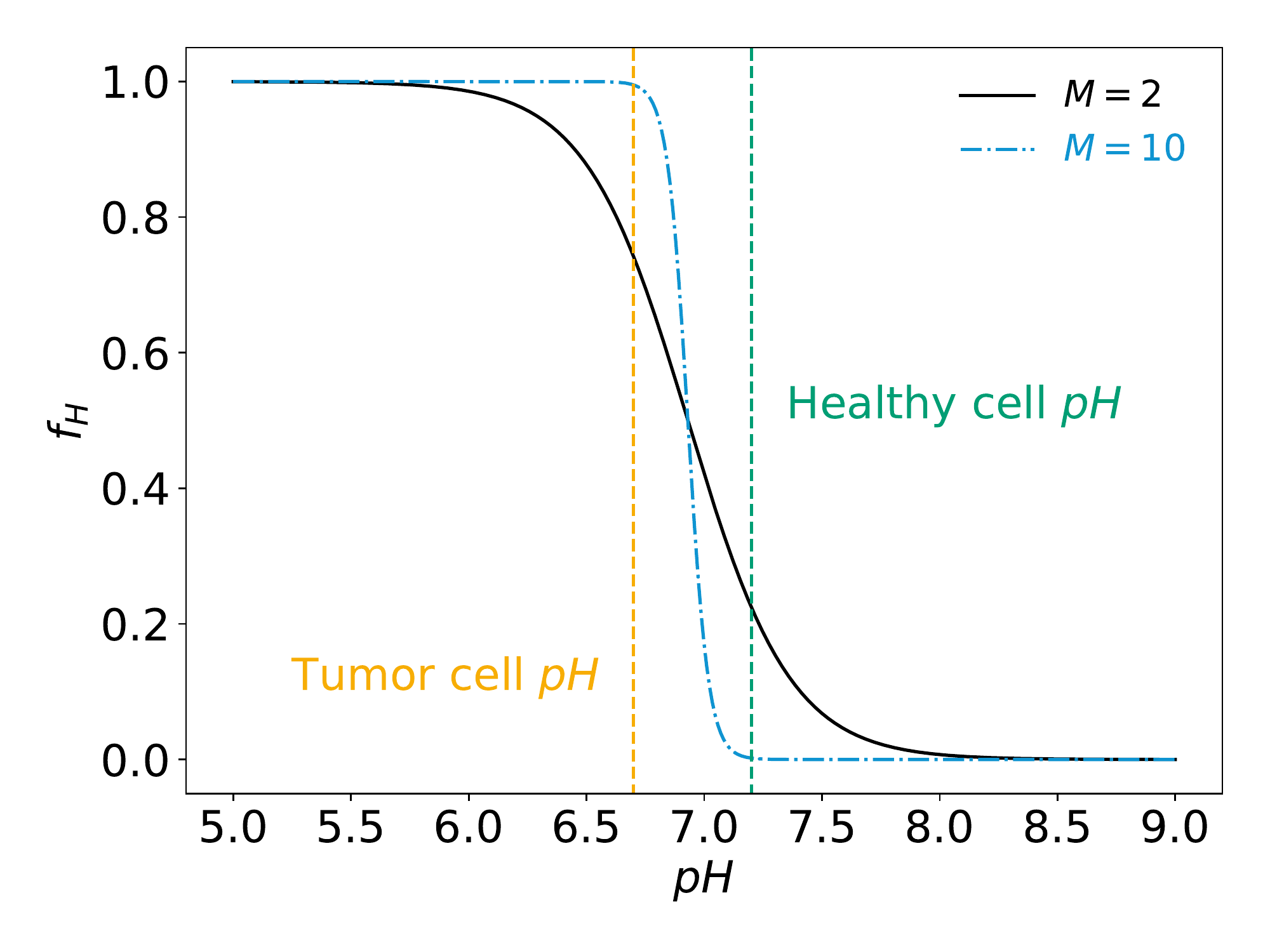}
	\caption{Schematic illustration of specific targeting of tumor cells by HPE. Vertical lines show the extracellular $pH$ of the healthy (green) and tumor (yellow) cells. $f_H$ curve calculated using Eq.~\ref{eq:fraction} (main text). Parameters: $\beta g_H = 5.6$, $pK_a = 4.5$ where $G_H=g_HM_H$, $M=M_H$.}
	\label{fig:permeation}
\end{figure}
\\
The observations on the eukaryotic (red blood) cells, together with our predictions of the width of the $pH$-mediated transitions, opens up potential applications in specific targeting of tumor cells by their small yet significantly lower extracellular $pH$. A possible scenario is depicted in Fig.~\ref{fig:permeation}. There we plotted the fraction of HPE in the hydrophobic state ($f_H$), which in this case corresponds to the interbilayer spacing of the lipid plasma membrane. The situation is analogous to disk formation; in that case $f_H$ should be replaced by the fraction of HPE in the disk state $f_D$ (Eq.~(\ref{SI_ksid})). The approximate extracellular $pH$ values of healthy cells and tumor cells are indicated by the green and red vertical lines in the Figure.
As can be seen in Fig.~\ref{fig:permeation}, with a degree of cooperativity of $M=2$, the HPE does not specifically target the tumor cells as the $pH$ range of the transition is roughly in between 6.5 and 7.5. The Figure implies that with $M=2$, HPE will have a preference for the interbilayer spacing of tumor cells, but healthy cells also are significantly affected. The situation is much more favorable (for healthy cells) if $M=10$: there, the transition is much sharper and HPE will only affect tumor cells. We stress here that $M$ is an increasing function of the number of ionic groups of the HPE but not equal to that number, due to Coulomb interactions and possibly other factors. In principle, both permeation (without the formation of disks) and disk formation are potential routes to target tumor cells by their relatively low extracellular $pH$. We briefly discuss the possible role of HPE architecture in disk formation and permeation below (\hyperref[SI_architecture]{\emph{SI: HPE architecture}}). Disk formation is expected to kill tumor cells directly. In the case of permeation of the HPE into the cell membrane, it is expected that HPE will move from the (low) extracellular $pH$ to the interbilayer spacing in the plasma membrane. From there, however, HPE are expected to move on to the cytoplasm of the tumor cell, which has a relatively high $pH$ compared to healty cells \cite{Sun2019}. In principle it should be possible to link low-molecular weight drug molecules onto HPE with chemical bonds that are broken once the polymers enter the cell cytoplasm. One such possibility is the use of disulfide bonds that are spontaneously broken in the reducing intracellular environment, see, for example, \cite{Yang2006}. As relatively high concentrations of HPE are required to cause significant membrane solubilization (1:1.25 lipid membrane to polymer weight ratio \cite{Zhang2015}), in practice HPE should be administered close to the expected tumor tissue. We expect that excess HPE is rapidly diluted and thereby harmless even if it is able to reach other tissue with low $pH$, such as the gastrointestinal tract \cite{Bazban-Shotorbani2017}.\\
\subsection*{HPE architecture}\label{SI_architecture}
\subsubsection*{Possible role of HPE architecture in disk formation and permeation}
\begin{table}[h!]
     \begin{center}
     \begin{tabular}{c c}
     \toprule
      Name & Chemical structure \\ 
      \midrule
      poly(alkyl acrylic acid) &  \includegraphics[scale=0.12, valign=c]{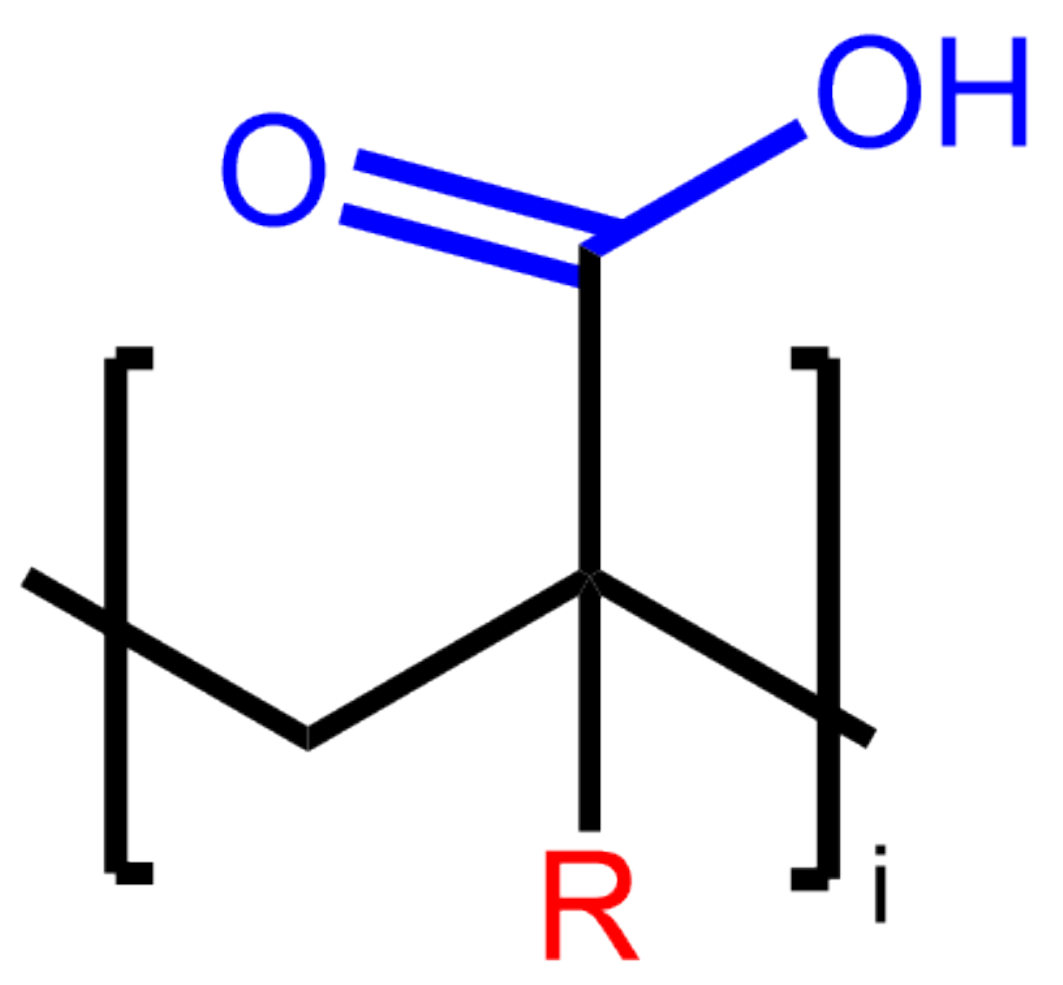} \\
      Styrene-maleic acid & \includegraphics[scale=0.12, valign=c]{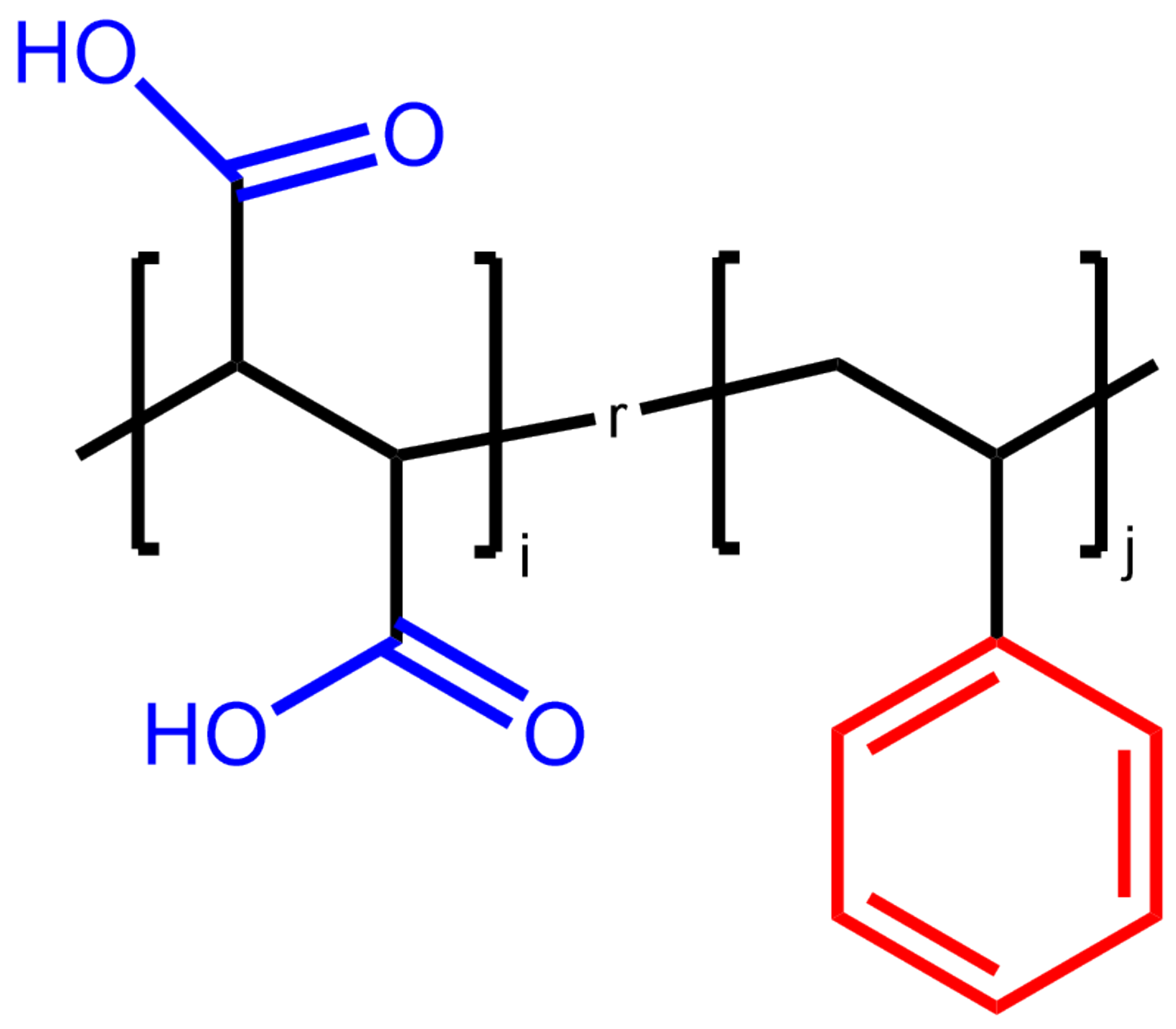} \\
      poly(2-(diakyl amino) ethyl methacrylate) & \includegraphics[scale=0.12, valign=c]{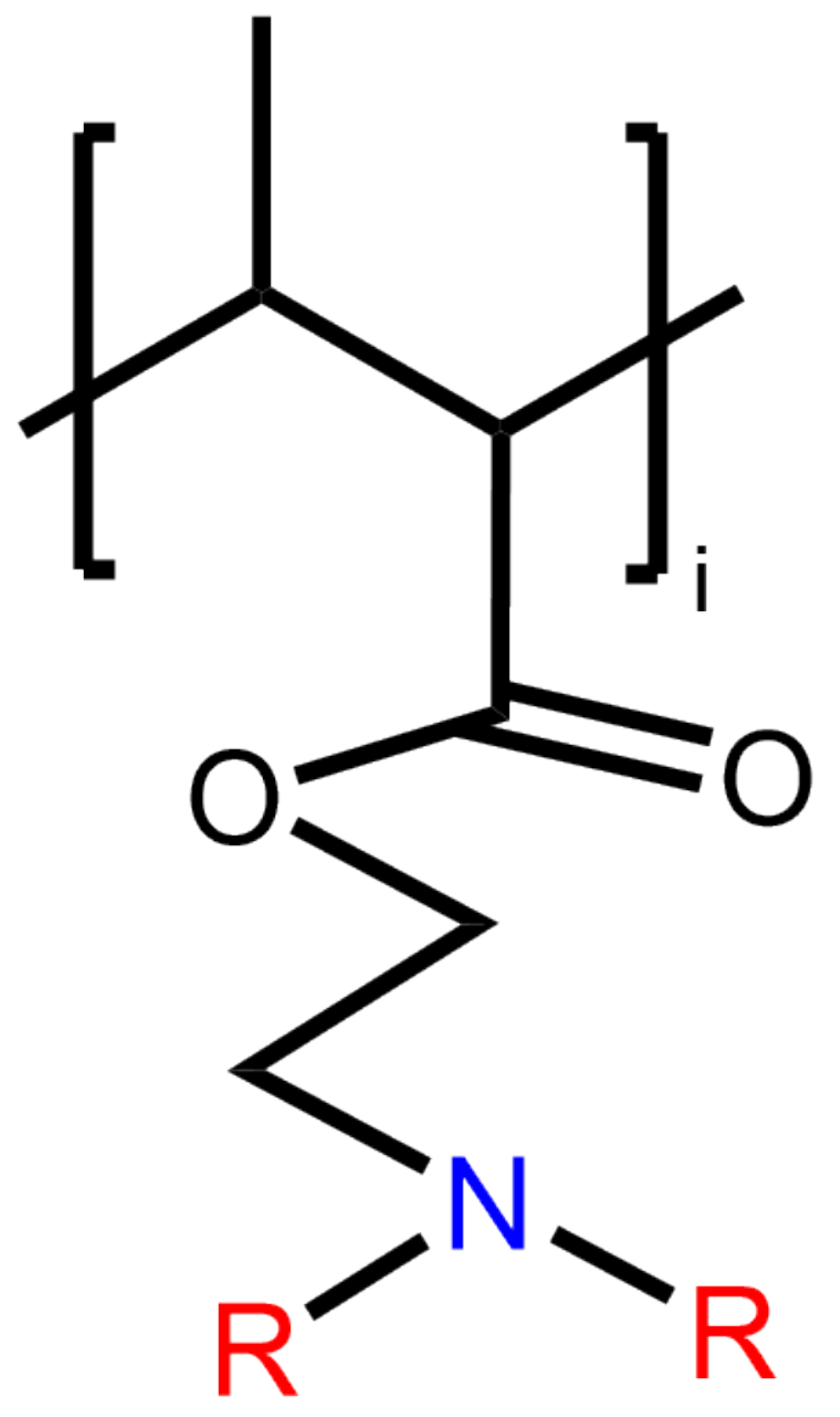} \\
      poly(carboxyalkyl acrylate) & \includegraphics[scale=0.12, valign=c]{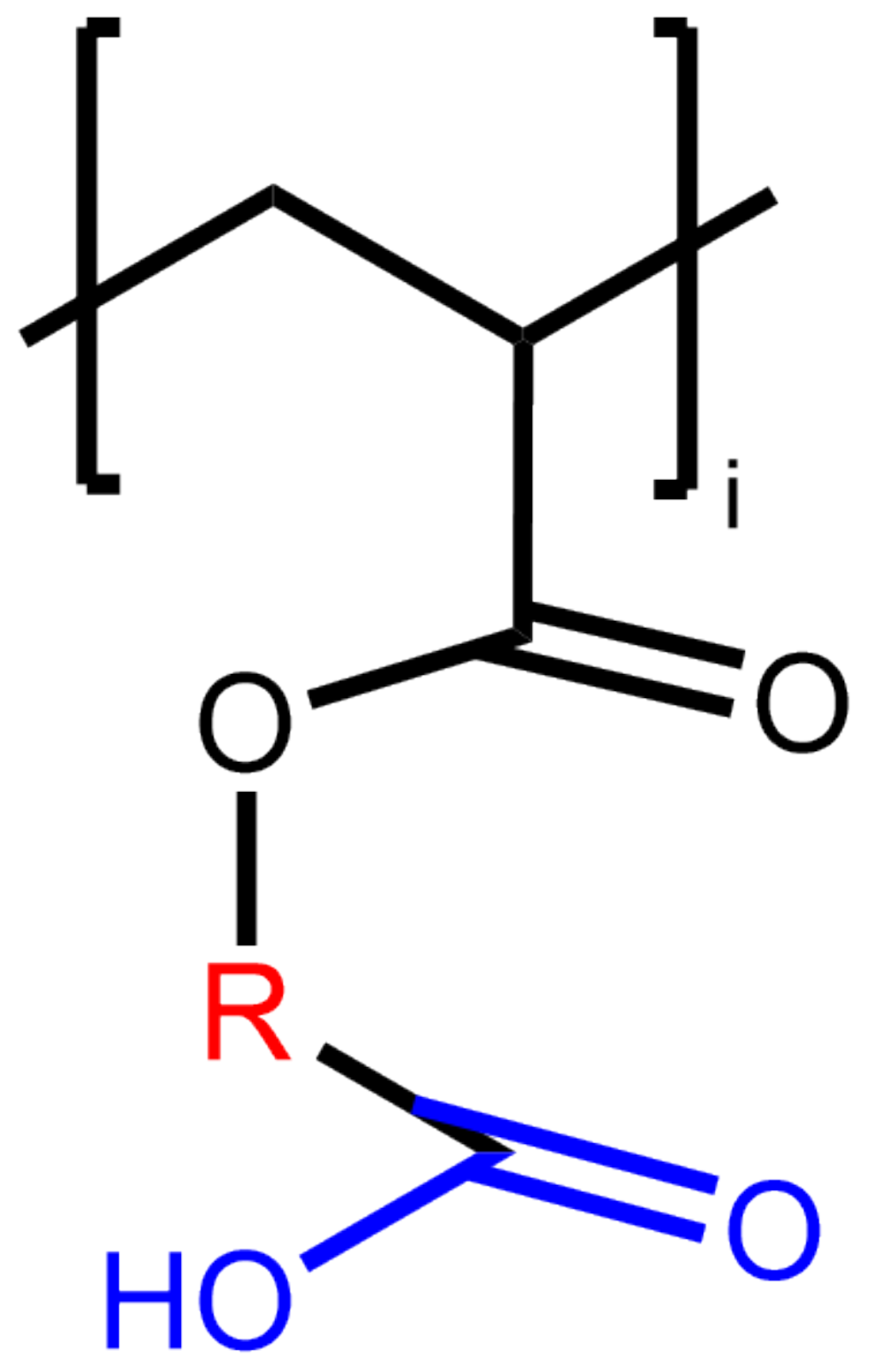} \\
       \bottomrule
      \end{tabular}
      \caption{General chemical structures of some hydrophobic polyelectrolytes. Hydrophobic groups (including R as alkyl chains) are depicted in red and ionizable groups are depicted in blue.}
      \label{table:polymer_structures}
      \end{center}
\end{table}
\noindent
Poly(carboxypentyl acrylate) (PCPA) has been observed to permeate DOPC (1,2-Dioleoyl-sn-glycero-3-phosphocholine) bilayers without any indication of disk formation \cite{Brodszkij2019}. When the $pH$ is decreased, PCPA has been shown to move to the interbilayer regions of the bilayer vesicles. Upon increasing $pH$ again, the polymer moves back to the aqueous state, where it effectively passes the bilayer membrane that separates the extra- and intra vesicle regions \cite{Brodszkij2019}. While the nature of the lipids may play a role, that is, contrary to the lipids in \cite{Thomas1994, Scheidelaar2016}, DOPC contains unsaturated bonds, our hypothesis is that its architecture makes PCPA unable to stabilize disks (see Table.~\ref{table:polymer_structures}). This is because in PCPA, the hydrophobic and carboxyl groups are combined in single side groups. The transition from an aqueous state to a state where the HPE are dissolved in the interbilayer regions is expected to be well described by the two-state model, Eqs. (\ref{eq:fraction}, \ref{theta}) in the main text. As the permeation transition for PCPA takes place around $pH = 5$, longer aliphatic chains will be required to tune the $pH$ to around neutral values. Likely, a polymer with mixed aliphatic chain lengths is necessary, similar to MAA-s-EAA in \cite{Thomas1994}, in order to sharply tune the transition $pH$.\\
\subsection*{HPE aggregation}\label{SI_aggregation}
\subsubsection*{Aggregation transition of poly(styrene-maleic acid) as a function of pH}
Aggregation and macroscopic precipitation of hydrophobic polyelectrolytes could be envisaged as a macroscopic consequence of the coil to globule transition and it may be expected that the transition pH values are similar.\\
\begin{figure}[h!]
	\centering
	\subfloat[\centering Fit of the solubility-insolubility transition (closed symbols) for various SMA variants, using Eq.~(\ref{eq:fraction}) (main text). Local parameters: $M=12.0,4.9,5.6$ for the $\frac{M_H}{M}=$ 2,3 and 4 variants respectively. Global parameters: $pK_a=2.4$ and $g_H=2.1k_BT$. Eq.~(\ref{pHagg}) was used to calculate $G_H$ for the different $\frac{M_H}{M}~(=\frac{j}{i})$ values. Open symbols represent titration ($\theta$) data. Note the general decorrelation between the titration and the $f_{H}$ data.]{{\includegraphics[width=0.5\linewidth]{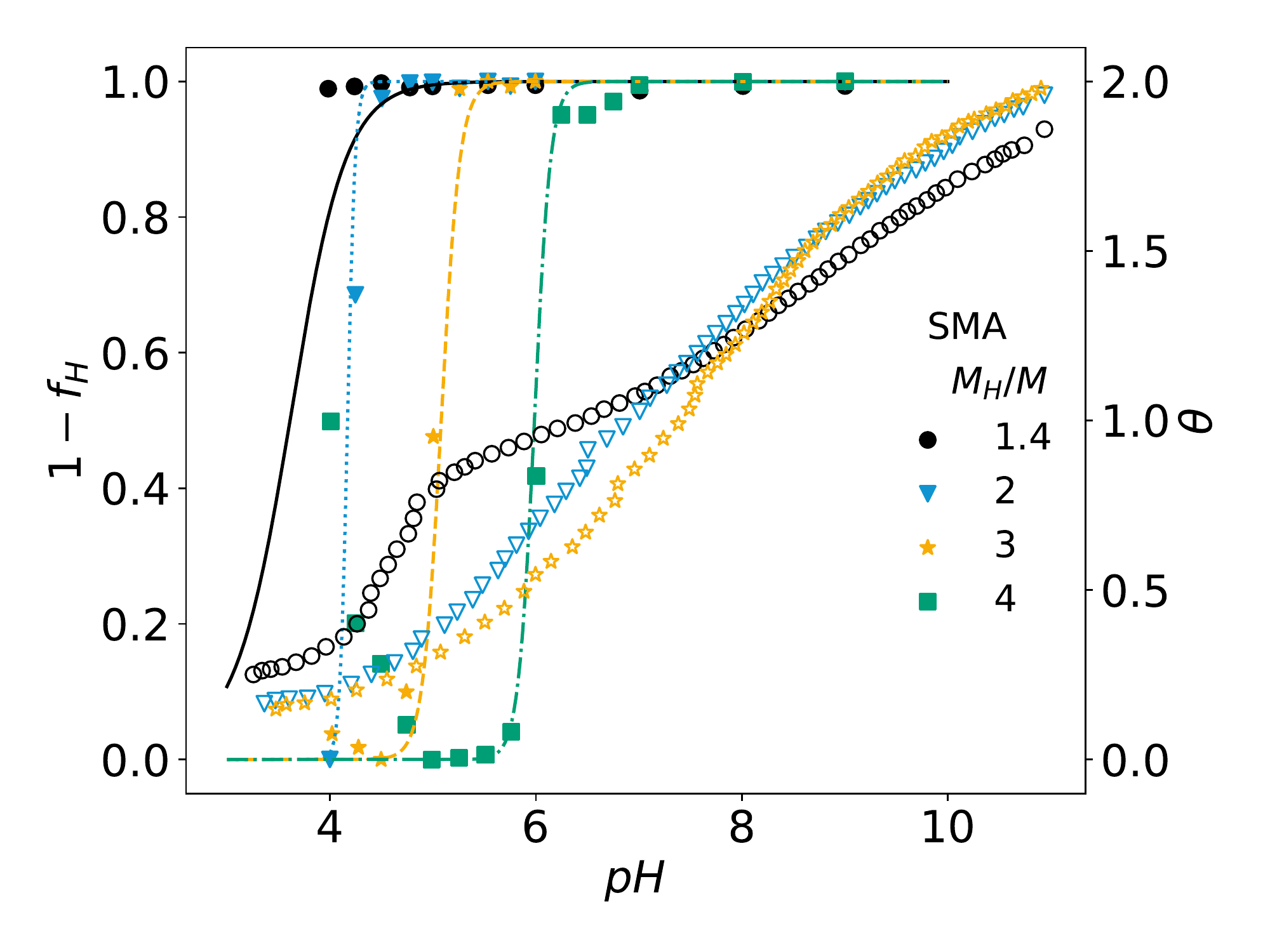} }}%
	\subfloat[\centering Transition $pH$ fit using Eq.~ (\ref{pHagg}).]{{\includegraphics[width=0.5\linewidth]{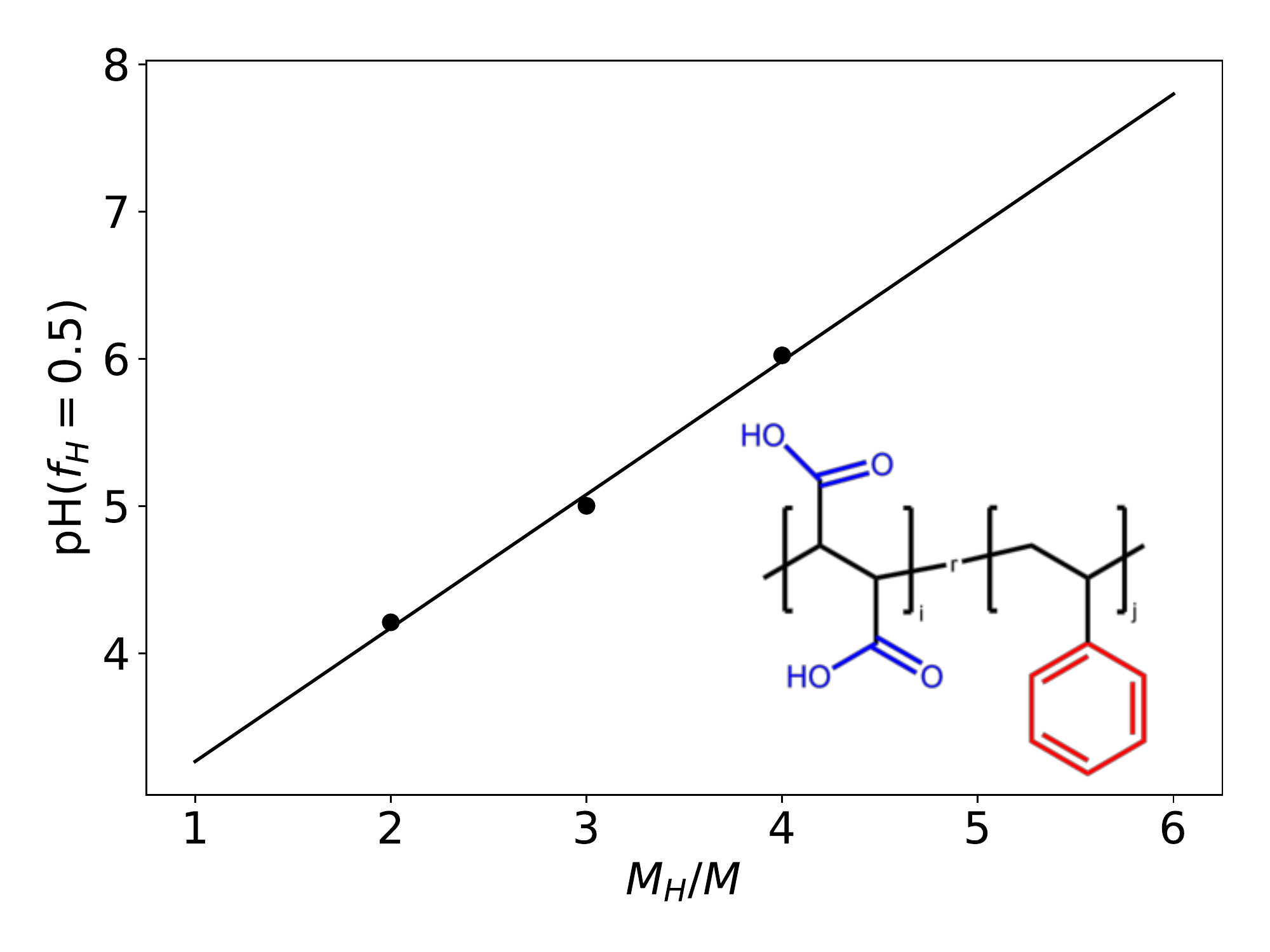} }}%
	\caption{Analysis of SMA aggregation data from Ref. \cite{Scheidelaar2016} (a) Turbidity and titration measurements for the SMA variants.(b) Turbidity transition $pH$ trend. Inset: Chemical structure of the polymer.} 
	\label{SMA1_fig}
\end{figure}
\\
Scheidelaar et al. investigate, using turbidimetry, the aggregation of a series of styrene-maleic acid polyelectrolytes with different compositions ($\frac{M_H}{M}=\frac{j}{i}$ in the inset of Fig.~\ref{SMA1_fig}b) as a function of the $pH$ of the solution \cite{Scheidelaar2016}. The data is reproduced in Fig.~\ref{SMA1_fig}a. As can be seen in Fig.~\ref{SMA1_fig}b, they observe a clear trend between the transition $pH$ and the styrene to maleic acid ratio that defines the hydrophobicity of these polymers.\\
The $pH$ where the transition takes place is where $\Xi_{aq} = \Xi_H = 1$ with $\Xi_{aq}$ given by Eq.~(\ref{eq:fraction}) in the main text. Assuming that in Eq.~(\ref{eq:fraction}) $G_H = g_H M_H$ with $M_H$ the number of hydrophobic styrene groups in the polymer, this $pH$ is given by 
\begin{equation}
	pH_{\text{aggr}} = pK_a + \text{0.4343}\beta g_H \frac{M_H}{M} \label{pHagg}.
\end{equation}
The results are shown in Fig.~\ref{SMA1_fig}b. The expected linear trend is present and a value of 2.1 is found for $\beta g_H$. An effective value of the $pK_a$ for the (first ionization of) the maleic acid groups can be found from the intercept of the fit and yields a value of 2.4. However as mentioned in the \hyperref[Membrane_section]{\emph{Membrane solubilization by disk formation}} section in the main manuscript and apparent from Fig.~\ref{SMA1_fig}a the ionization state of the polymers and their macroscopic behavior seems to be uncorrelated and therefore this analysis does not yield meaningful values.\\
Fig.~\ref{SMA1_fig}a shows a fit of the solubility-insolubility transition using the value of $g_H = 2.1$ as derived from Fig.~\ref{SMA1_fig}b. Ignoring the 1.4:1 transition where a sharpness cannot be ascertained due to a lack of data, the fitted values of $M$ are, compared to the coil to globule transitions described earlier, fairly large. Although we may be inclined to postulate that this a consequence of a much stronger adherence to a two- state, well defined, transition, an analysis of the titration of these polymers paints a different picture. The $\theta$ curve seems to be mostly uncorrelated from the measured aggregation transition. Therefore, a more likely scenario is that the chains go through a relatively broad coil to globule transition at higher pH values and once the globules have a low enough charge they can aggregate and precipitate out. This interpretation has been illustrated in Fig.~\ref{fig:aggregation_diag}. Supporting this hypothesis is the fact that aggregation occurs at a higher $pH$ when the salt concentration is increased \cite{Scheidelaar2016}, due to the increased screening allowing for closer packing of the charges.\\
\begin{figure}[h]
\centering
\includegraphics*[width=1\linewidth]{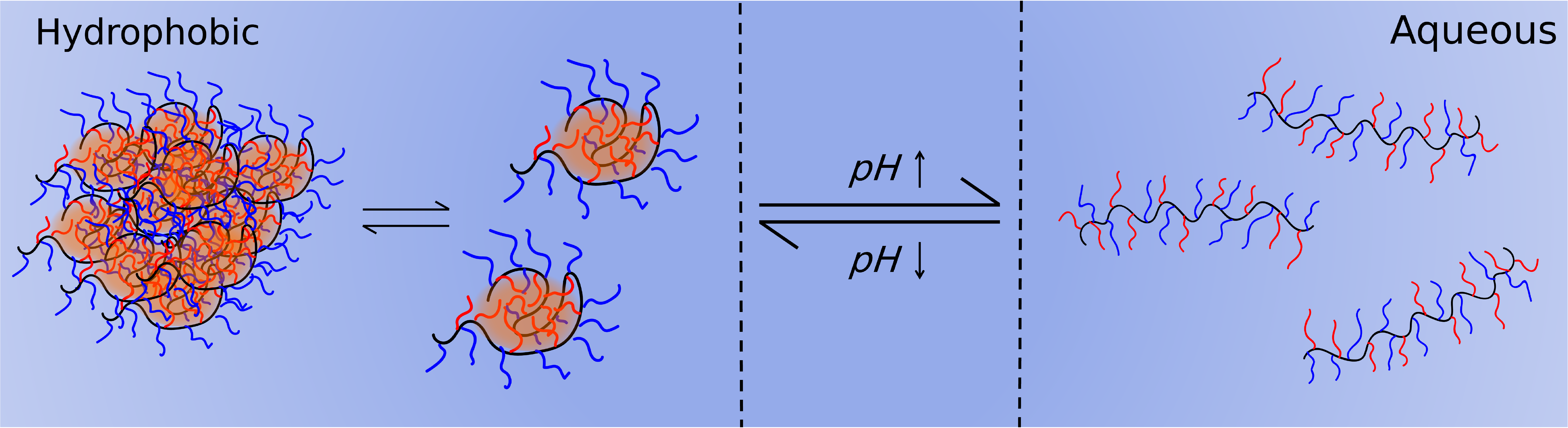}
\caption{Schematic illustration of the proposed aggregation process. A $pH$-dependent coil to globule transition is followed by the aggregation of the globules into macroscopic aggregates. Red and blue polyelectrolyte side groups represent hydrophobic and hydrophilic (acidic) molecular moieties, respectively. Orange shading highlights the hydrophobic reservoirs that stabilise these hydrophobic side groups.}
\label{fig:aggregation_diag}
\end{figure}
\subsection*{Influence of intermediate states on transition broadening}\label{SI_broadening}
\begin{figure}
    \centering
	\subfloat[\centering $f_{exp}$ parameters: $M=40$, $M_{int}=36$, $g_H=4.5$, $pK_{a}=4.5$, $g_{int}=4$ and $pK_{a,int}=4.5$. Where $G_H=g_HM$ and $G^*_{int}=g_{int}M$. $\overline{f_{H,fit}}$ parameters: $\overline{M}=4.2$, $pK_a=4.5$ and $\overline{g_H}=11.9$, where $\overline{G_H}= \overline{g_HM}$. ]{{\includegraphics[width=0.45\linewidth]{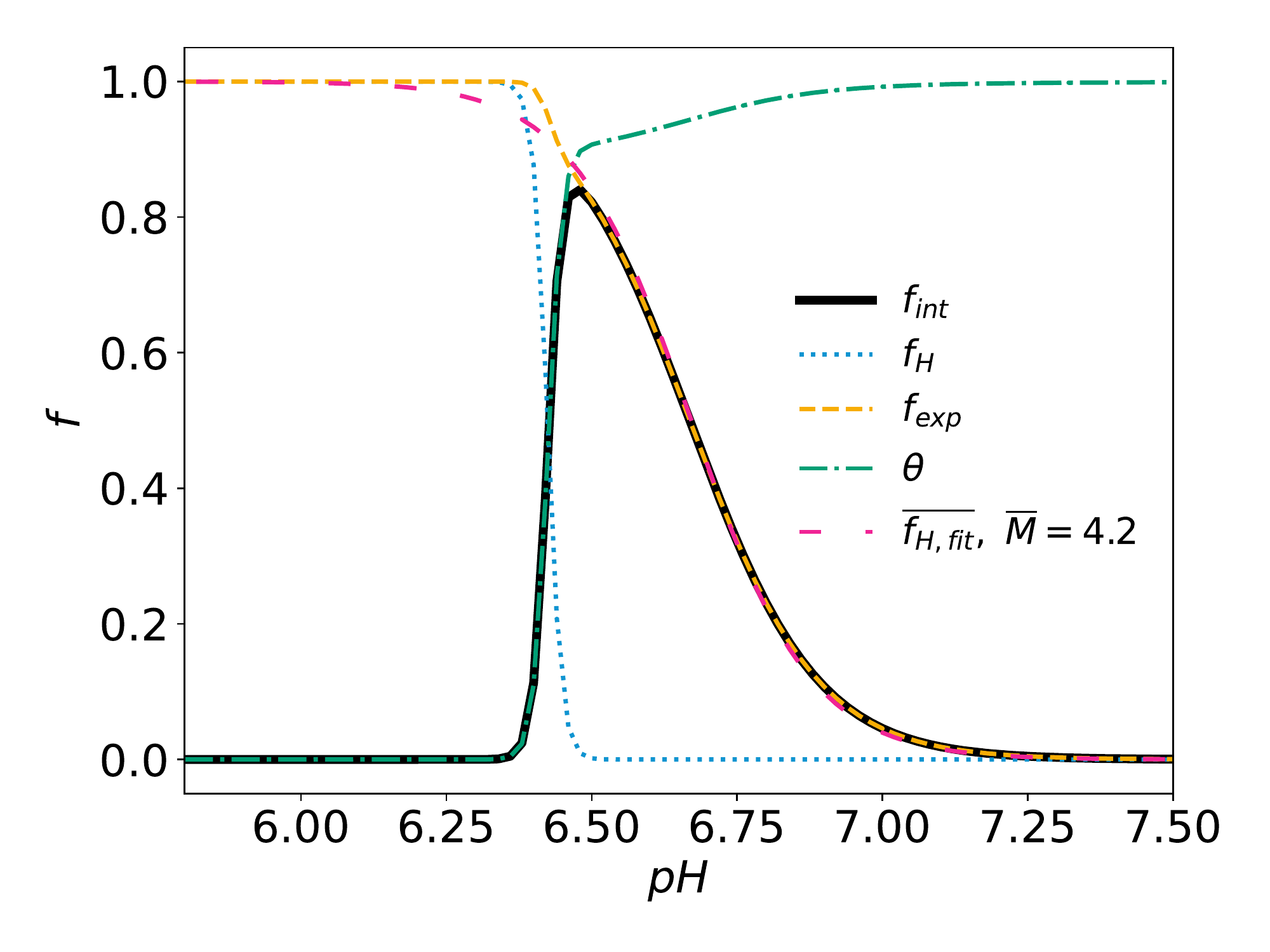} }}%
	\subfloat[\centering $f_{exp}$ parameters: $M=40$, $M_{int}=20$, $g_H=4.5$, $pK_{a}=4.5$, $g_{int}=2$ and $pK_{a,int}=4.5$. Where $G_H=g_HM$ and $G^*_{int}=g_{int}M$. $\overline{f_{H,fit}}$ parameters: $\overline{M}=19.9$, $pK_a=4.5$ and $\overline{g_H}=11.9$, where $\overline{G_H}= \overline{g_HM}$.]{{\includegraphics[width=0.45\linewidth]{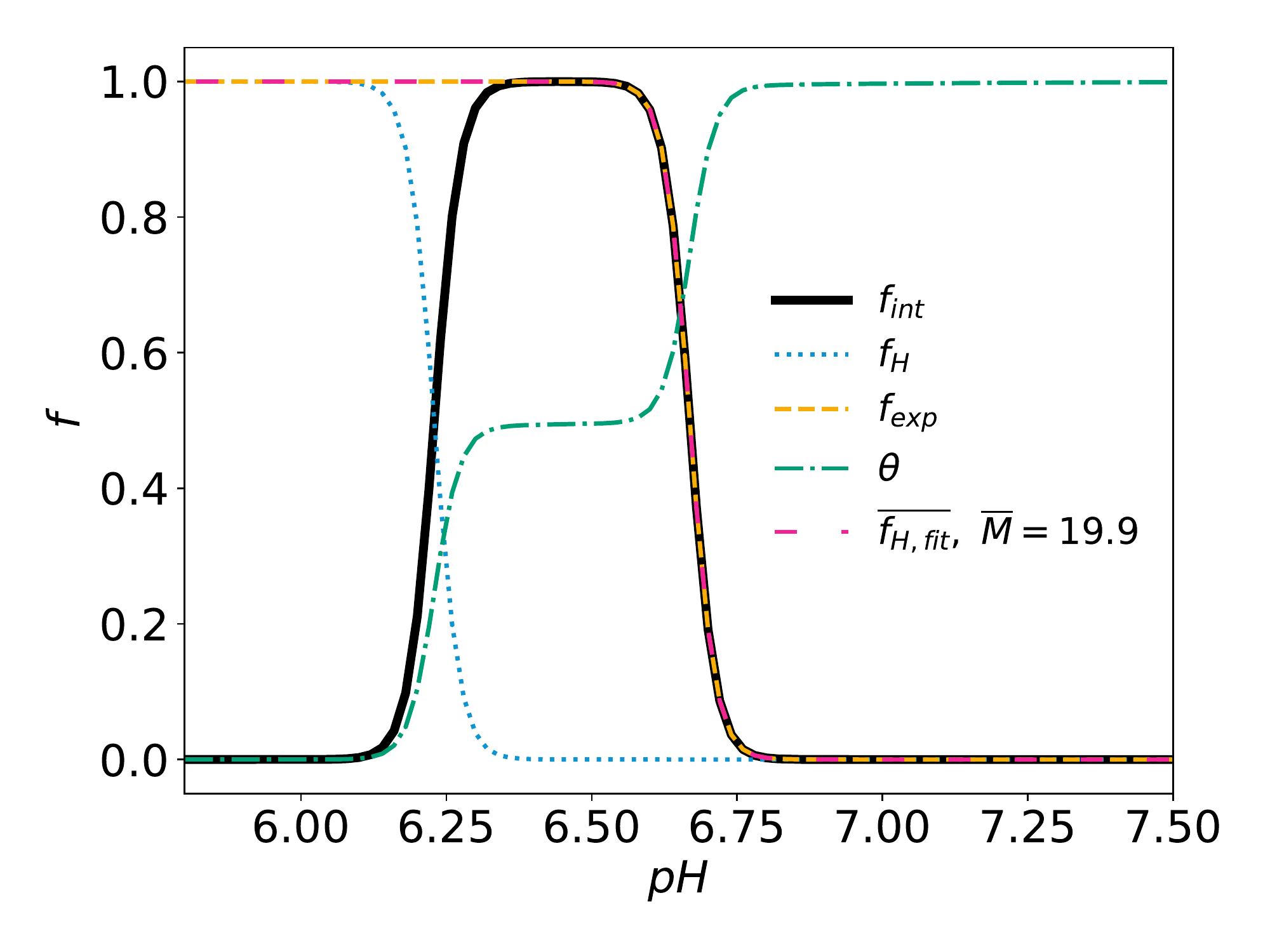} }}%
	\caption{Calculated transitions, $f_{exp}=f_{H}+f_{int}$ (Eq.~(\ref{f_int})), for an acidic HPE with a third intermediate state between the aqueous and hydrophobic state. The equation for the hydrophobic fraction in a two-state system , $\overline{f_{H,fit}}$ (Eq.~(\ref{eff_frac})), with a cooperativity parameter $\overline{M}$ is then fitted to $f_{exp}$ to find an effective value for the cooperativity. The ionization fraction, $\theta$ (Eq.~(\ref{theta_int})), is also shown and presents, in both plots, deviations from the usual sigmoidal shape seen for two-state systems. Depending on the relative values of $G^*_{int}$ and $M_{int}$ we may get a narrow range (a) or a wide range (b) of stability for the intermediate state. In this figure we have assumed a polymer with equal numbers of ionizable, $M$, and hydrophobic side groups, $M_H$, while in the aqueous state. Although $M_{int}$, the number of ionizable groups on the chain in the intermediate state, will differ from $M$, the number of hydrophobic groups will not change regardless of the state of the polymer. Therefore both of the hydrophobic parameters, $g_H$ and $g_{int}$, are defined with respect to $M$.} 
	\label{interstatefig}
\end{figure}

A core assumption of the MWC model is that there are well-defined conformational states of the substrates that lead to well-defined transitions. The sharpest transitions, in the case of HPE, occur when there are only two different conformations with different hydrophobic penalties. An obvious source of broadening may be the presence of intermediate states between the two extremes. To illustrate this we take only one intermediate state with hydrophobic penalty $G_{int}^*$ and number of ionizable groups $M_{int}$. We also set the $pK_a$ in each state as equal. There, the situation is comparable to the presence of disks (Eq.~(\ref{SI_ksid})) as described in the previous section,
\begin{linenomath}
\begin{equation}
	\Xi_{int} =\exp{(-\beta G_{int}^*)} \left(1 + 10^{pH-pK_a} \right)^{M_{int}}.   \label{kinter}
\end{equation}
\end{linenomath}
The quantity experimentally measured to represent the hydrophobic or aqueous state now becomes relevant. In Fig.~\ref{interstatefig}, we present a scenario where the measured quantity (for example, scattering intensity) responds equally to both the hydrophobic and intermediate states. In combination with the expressions for the aqueous, $\Xi_{aq}$ and hydrophobic, $\Xi_{H}$, partition functions Eqs.~(\ref{ksiH},\ref{ksiaqac}), the system can be described as follows
\begin{linenomath}
\begin{align}
    \Xi=\Xi_{aq}&+\Xi_{H}+\Xi_{int}\\
    f_{exp}=\frac{\Xi_{int}}{\Xi} &+ \frac{\Xi_{H}}{\Xi} =f_{int}+f_{H}. \label{f_int}   
\end{align}
\end{linenomath}
$f_{exp}$ is the experimentally measured quantity composed of the sum of the hydrophobic and intermediate state fractions. 
The values chosen for $M_{int}$ and $G^*_{int}$ are important to make sure the intermediate phase is stable at intermediate $pH$ values between the aqueous and hydrophobic states. As shown in the main text for disks (Eqs.~(\ref{pHda},\ref{pHhd})), the transition $pH$ values for the hydrophobic-intermediate and intermediate-aqueous states for an acidic HPE are as follows:
\begin{linenomath}
\begin{align}
    pH_{H-inter} &= pK_a + \text{0.4343}\beta \frac{G^*_{int}}{M_{int}}\\
    pH_{inter-aq} &= pK_a + \text{0.4343}\beta \frac{G_H-G^*_{int}}{M-M_{int}}.\label{f_int_trans}   
\end{align}
\end{linenomath}
The stability condition for the intermediate phase, assuming $0<M_{inter}<M$ is $pH_{inter-aq}>pH_{H-int}$. This leads to the following condition:
\begin{equation}
    \frac{G_H}{G^*_{int}}>\frac{M_{H}}{M_{int}}.
\end{equation}
The particular values chosen for these parameters determine the range of $pH$ values that the intermediate phase is stable for. In practice very low values of $G^*_{int}$ will push $pH_{inter-aq}$ to unphysically high values. In Fig.~\ref{interstatefig} we plot two examples of a system with an intermediate state. $f_{exp}$ is then fitted to a two-state model, $\overline{f_{H,fit}}$, based on Eqs.~(\ref{ksiH},\ref{ksiaqac})), to assess the apparent broadening of the transition
\begin{linenomath}
\begin{equation}
    \overline{f_{H,fit}} = \left( \exp{ (-\beta \overline{G_{H}})} \left( 1 + 10^{pH-pK_a} \right)^{\overline{M}} \right)^{-1} \label{eff_frac}.
\end{equation}    
\end{linenomath}
 $\overline{M}$ will be the effective cooperativity parameter for the transition and $\overline{G_H}$ the effective hydrophobic penalty. The values of $\overline{M}$ for both plots in Fig.~\ref{interstatefig} are substantially smaller than $M$ and it is clear that the broadness of the transitions correlates with the value of $M-M_{int}$. This illustrates how the presence of an intermediate state may change the sharpness of a transition from the value of $M$ predicted from the structure of the polymer. Moreover, as seen in Fig.~\ref{interstatefig}b, the shape of the $f_{exp}$ curve derived from a three-state system may be indistinguishable from a two-state system transition. The transitions also present a higher effective transition $pH$ that leads to a larger value for the fitted hydrophobic penalty, $\overline{g_H}$ where $\overline{G_H}= \overline{g_HM}$, when compared to the $g_H$ ($G_H=g_HM$) values used to calculate the transitions.\\

 The ionization fraction, $\theta$, was also investigated for the 3-state system. Using the general expression from Eq.~(\ref{SI_theta}), the expression for $\theta$ in a three-state system is
\begin{linenomath}
\begin{equation}
    \theta=\frac{10^{pH-pK_a}}{1+10^{pH-pK_a}}\frac{1}{\Xi}\left(\exp{(-\beta G_{H})} \left(1 + 10^{pH-pK_a} \right)^{M}+\frac{M_{int}}{M}\exp{(-\beta G_{int}^*)} \left(1 + 10^{pH-pK_a} \right)^{M_{int}}\right). \label{theta_int}
\end{equation}
\end{linenomath}
This expression is analogous to Eq.~(\ref{thetaD}) in the main text and presents the $\frac{10^{pH-pK_a}}{1+10^{pH-pK_a}}$ term that is discussed in the \hyperref[micelles_section]{\emph{Diblock Micelles}} section of the main text. The $\theta$ curves in Fig.~\ref{interstatefig}, in particular part (b), present substantial deviations from the usual sigmodial shapes seen for two-states systems and most of the experimental data shown in the main text. This is due to the difference in the number of ionizable groups between the intermediate and aqueous states which results in a function with two distinct steps. Inclusion of a larger number of intermediate states would smooth out the curves for $\theta$ and match the experimental data more effectively.\\
Coil-globule transitions are known to have many intermediate steps between the most extended and most coiled up states \cite{Ulrich2005,Kiriy2002}. An extension of what is laid out above could be made to incorporate more intermediate states, which would also transform the $\theta$ curves into more physical shapes. Specific knowledge of the functions describing $G^*_{int}$ and $M_{int}$ for these intermediate states will be required to present a quantitative description of the coil-globule transition.\\
\bibliographystyle{pnas-new.bst}
\bibliography{HPE}
\end{document}